\documentstyle[11pt]{article}
\begin{document}

\title{\textbf{Generalized second law of thermodynamics on the apparent horizon in
modified Gauss-Bonnet gravity}}

\author{A. Abdolmaleki$^{1}$\thanks{AAbdolmaleki@uok.ac.ir} ,
T. Najafi$^{2}$\thanks{t.najafi90@gmail.com} ,\\
$^{1}$\small{Center for Excellence in Astronomy \& Astrophysics of
Iran (CEAAI-RIAAM), Maragha, Iran}\\$^{2}$\small{Department of
Physics, University of Kurdistan, Pasdaran St., Sanandaj, Iran} }

\maketitle

\begin{abstract}
Modified gravity and generalized second law (GSL) of thermodynamics
are interesting topics in the modern cosmology. In this regard, we
investigate the GSL of gravitational thermodynamics in the framework
of modified Gauss-Bonnet gravity or $f(G)$-gravity. We consider a
spatially FRW universe filled with the matter and radiation enclosed
by the dynamical apparent horizon with the Hawking temperature. For
two viable $f(G)$ models, we first numerically solve the set of
differential equations governing the dynamics of $f(G)$-gravity.
Then, we obtain the evolutions of the Hubble parameter, the
Gauss-Bonnet curvature invariant term, the density and equation of
state parameters as well as the deceleration parameter. In addition,
we check the energy conditions for both models and finally examine
the validity of the GSL. For the selected $f(G)$ models, we conclude
that both models have a stable de Sitter attractor. The equation of
state parameters behave quite similar to those of the $\Lambda$CDM
model in the radiation/matter dominated epochs, then they enter the
phantom region before reaching the de Sitter attractor with
$\omega=-1$. The deceleration parameter starts from the
radiation/matter dominated eras, then transits from a cosmic
deceleration to acceleration and finally approaches a de Sitter
regime at late times, as expected. Furthermore, the GSL is respected
for both models during the standard radiation/matter dominated
epochs. Thereafter when the universe becomes accelerating, the GSL
is violated in some ranges of scale factor. At late times, the
evolution of the GSL predicts an adiabatic behavior for the
accelerated expansion of the universe.
\end{abstract}


\clearpage

\section{Introduction}

Various cosmological observations, coming from the type Ia
supernovae (SNeIa) surveys \cite{SN}, the large scale structure
(LSS) \cite{LS}, the cosmic microwave background (CMB) anisotropy
spectrum \cite{CMB1, CMB2} and the Hubble parameter $H(z)$ \cite{H},
have indicated that the universe is in a phase of accelerated
expansion. Regarding the accelerated expansion of the universe,
there are two main categories of probable solutions. One is to
assume that in the context of general relativity (GR), the universe
is dominated by a new cosmic fluid with negative pressure. This kind
of exotic matter which violates the strong energy condition is so
called ``dark energy`` (DE) \cite{Padmanabhan,Sahni}. Another
alternative, originates from the modification of gravity \cite{RoG}.
In modified gravity (MG) theories, there is no require for DE with
exclusive properties, but instead, the action contains a general
function of invariants obtained from the Riemann curvature tensor
such as the Ricci scalar, $R$, \cite{fR} or the Gauss-Bonnet
invariant term, $G$, \cite{fG} or the torsion scalar, $T$ \cite{fT}.
Moreover, in \cite{Sobouti} it was shown that MG may serve as dark
matter (DM).

One of interesting alternative theories of gravity is modified
Gauss-Bonnet gravity, so-called $f(G)$-gravity, where $f(G)$ is a
general function of the Gauss-Bonnet curvature invariant term
$G=R^{2}-4R_{\mu\nu}R^{\mu\nu}+R_{\mu\nu\rho\sigma}R^{\mu\nu\rho\sigma}$
\cite{fG,Barrow0,JCAP,PLB,Living,Felic,de Sitter}. The
$f(G)$-gravity can justify the present accelerated expansion of the
universe without resorting to DE. Besides, it can also describe the
phantom divide line crossing as well as the cosmic transition from
deceleration to acceleration phase \cite{Wald,AG,Metsaev}. The
cosmologically viable $f(G)$ models need to be close to the
$\Lambda$CDM model in the deep matter era, but the deviation from it
becomes important at late times on cosmological scales. An
appreciable deviation from the $\Lambda$CDM cosmology yields the
modification of the matter power spectrum, which can be used as a
crucial tool to distinguish $f(G)$-gravity from the $\Lambda$CDM
model.

Additionally, the thermodynamical interpretation of gravity is one
of another interesting topics in modern cosmology. From the
viewpoint of the physics of the black holes, there is a deep rooted
connection between thermodynamics and gravity \cite{BekHaw}. This
connection was first discovered in the Einstein gravity for the 
Rindler spacetime \cite{Jacobson}. It was also shown that by
assuming the geometric entropy given by a quarter of the apparent
horizon area of a Friedmann-Robertson-Walker (FRW) universe, the
Friedmann equation in the Einstein gravity can be written in the
form of the first law of thermodynamics \cite{Cai05}. The connection
between thermodynamics and gravity has also been investigated in
$f(R)$-gravity and scalar-tensor theory \cite{Akbar12},
$f(T)$-gravity \cite{MK}, Lovelock theory \cite{Akbar} and
braneworld scenarios (such as DGP, RSI and RSII) \cite{Sheykhi1}.

It is also of great interest to generalize our discussion to study
the generalized second law (GSL) of thermodynamics. The GSL states
that the entropy of matter inside the horizon of the universe plus
the geometric entropy of the horizon is non-decreasing with time
\cite{Cai05}. Note that the ordinary second law of thermodynamics
only deals with the entropy of matter inside the universe. In the
Einstein gravity, it was shown that the GSL in the presence of DE is
always satisfied \cite{Izquierdo1}. The validity of the GSL was also
examined in different theories of MG \cite{Izquierdo1}-\cite{Geng}.
Here our main aim is to explore the GSL and the thermodynamics of
the apparent horizon in $f(G)$-gravity, and obtain the condition for
the GSL to be satisfied. The paper is structured as follows. In
section 2, we briefly review the $f(G)$-gravity. In section 3, we
investigate the GSL of thermodynamics on the dynamical apparent
horizon with the Hawking temperature. In section 4, we study the
dynamics of $f(G)$-gravity. In section 5, we give numerical results
obtained for the evolution of some cosmological parameters, check
the energy conditions and examine the validity of the GSL for two
viable $f(G)$ models. Section 6 is devoted to conclusions.

\section{THE $f(G)$ THEORY OF GRAVITY}

The action of modified Gauss-Bonnet gravity is given by \cite{JCAP}:
\begin{equation}\label{action}
 I=\int {\rm d}^4x\sqrt{-g}\left(\frac{1}{2k^2} R + f(G)+ L_{\rm r} + L_{\rm m}\right),\label{action1}
\end{equation}
where $k^2=8\pi G_{\rm N}=1$ and $G_{\rm N}$ is the Newtonian
gravitational constant. Also $g$, $R$, $L_{\rm r}$, $L_{\rm m}$ and
$f(G)$ are the determinant of metric $g_{\mu\nu}$, Ricci scalar, the
matter Lagrangian, the radiation Lagrangian and a general function
of the Gauss-Bonnet term, respectively. The Gauss-Bonnet curvature
invariant term is defined as
\begin{equation}
G=R^{2}-4R_{\mu\nu}R^{\mu\nu}+R_{\mu\nu\rho\sigma}R^{\mu\nu\rho\sigma}.\label{Gauss-Bonnet}
\end{equation}
Taking variation of the action (\ref{action1}) with respect to
$g_{\mu\nu}$ leads to the field equations
\begin{eqnarray}\label{field1}
T_{\mu\nu}= R_{\mu\nu} - \frac{1}{2}R g_{\mu\nu}+
8[R_{\mu\rho\nu\sigma} + R_{\rho\nu}g_{\sigma\mu}
-R_{\rho\sigma}g_{\nu\mu}  - R_{\mu\nu}g_{\sigma\rho}
+R_{\mu\sigma}g_{\upsilon\rho}\\\nonumber
+\frac{1}{2}R(g_{\mu\nu}g_{\sigma\rho}-
g_{\mu\sigma}g_{\nu\rho})]\nabla^{\rho} \nabla^{\sigma}f_{\rm G} +
(Gf_{\rm G}-f)g_{\mu\nu},
\end{eqnarray}
where $f_{\rm G}={\rm d}f/{\rm d}G$. Also $T_{\mu\nu}$ and
$\nabla^{\rho}$ are the energy-momentum tensor and the covariant
derivative operator, respectively. Now we consider a spatially flat
universe described by the FRW metric

\begin{equation}\label{FRW 1}
{\rm d}s^2 = -{\rm d}t^2+a^2(t)\Big({\rm d}x^2 + {\rm d}y^2 + {\rm
d}z^2 \Big),
\end{equation}
where $a(t)$ is the scale factor. Consequently, we have
\begin{equation}\label{R,G}
R=6\left(\dot{H}+2H^{2}\right), ~~~~
G=24H^{2}\left(\dot{H}+H^{2}\right),
\end{equation}
where $H=\dot{a}/a$ is the Hubble parameter and an overdot stands
for a derivative with respect to the cosmic time $t$. In terms of
the deceleration parameter $q=-1-\dot{H}/H^2$, Eq. (\ref{R,G}) can
be written as
\begin{equation}\label{R1,G1}
R=6H^{2}\left(1-q\right),~~~~ G=24H^{4}q.
\end{equation}

Substituting the FRW metric (\ref{FRW 1}) into the field equations
(\ref{field1}) yields the Friedmann equations in $f(G)$-gravity as
\begin{equation}\label{fr 1}
3H^{2}=Gf_{\rm G} - f -24H^{3}\dot{f}_{\rm G}+\rho_{\rm r}
+\rho_{\rm m},
\end{equation}

\begin{equation}\label{fr 2}
-2\dot{H}= -8H^{3}\dot{f}_{\rm G} +16H\dot{H}\dot{f}_{\rm G}
+8H^{2}\ddot{f}_{\rm G}+\frac{4}{3}\rho_{\rm r} +\rho_{\rm m},
\end{equation}
where $\rho_{\rm m}$ and $\rho_{\rm r}$ are the energy density of
matter and radiation, respectively.

The Friedmann equations (\ref{fr 1}) and (\ref{fr 2}) can be
rewritten in the standard form as \cite{Capozziello2}

\begin{equation}\label{frid1}
H^{2}=\frac{1}{3}\rho_{\rm t},\label{Fr}\label{Fr}
\end{equation}

\begin{equation}\label{frid2}
\dot{H}=-\frac{1}{2}\big(\rho_{\rm t}+p_{\rm t}\big),\label{Hdot}
\end{equation}
where $\rho_{\rm t}$ and $p_{\rm t}$ are the total energy density
and pressure defined as
\begin{eqnarray}
&&\rho_{\rm t}=\rho_{\rm m} +\rho_{\rm r} +\rho_{\rm G},\\
&&p_{\rm t}=p_{\rm m}+p_{\rm r}+p_{\rm G}.
\end{eqnarray}
Here $\rho_{\rm G}$ and $p_{\rm G}$ are the energy density and
pressure due to the $f(G)$ contribution defined as
\begin{equation}\label{ro G}
\rho_{\rm G}= Gf_{\rm G} - f -24H^{3}\dot{f}_{\rm G},
\end{equation}
\begin{equation}\label{p G}
p_{\rm G}=16H^{3}\dot{f}_{\rm G}+16H\dot{H}\dot{f}_{\rm
G}+8H^{2}\ddot{f}_{\rm G}-Gf_{\rm G}+f.
\end{equation}

By using of Eqs. (\ref{ro G}) and (\ref{p G}), one can obtain the
equation of state (EoS) parameter due to the $f(G)$ contribution as
\cite{JCAP}
\begin{equation}\label{w DE}
\omega_{\rm G}=\frac{p_{\rm G}}{\rho_{\rm G}}=
\frac{16H^{3}\dot{f}_{\rm G}+16H\dot{H}\dot{f}_{\rm
G}+8H^{2}\ddot{f}_{\rm G}-Gf_{\rm G}+f}{Gf_{\rm G}-f
-24H^{3}\dot{f}_{\rm G}}.
\end{equation}
Also from Eqs. (\ref{frid1}) and (\ref{frid2}), the effective EoS
parameter can be obtained as
\begin{equation}\label{w eff}
\omega_{\rm eff}=-1-\frac{2\dot{H}}{3H^{2}}.
\end{equation}
Moreover, the continuity equations governing the pressureless matter
($p_{\rm m}=0$), the radiation ($p_{\rm r}=\rho_{\rm r}/3$) and the
$f(G)$ contribution satisfy
\begin{equation}\label{co eq2}
\dot{\rho}_{\rm m}+3H\rho_{\rm m}=0,
\end{equation}

\begin{equation}\label{co eq1}
\dot{\rho}_{\rm r}+4H\rho_{\rm r}=0,
\end{equation}

\begin{equation}\label{co eq3}
\dot{\rho}_{\rm G}+3H(\rho_{\rm G}+ p_{\rm G})=0.
\end{equation}
Equations (\ref{fr 1}), (\ref{fr 2}), (\ref{co eq2}) and (\ref{co
eq1}) determine the dynamics of the $f(G)$-gravity system
(\ref{action}) in a homogeneous and isotropic background. We will
study this issue in section \ref{dynamics}.

\section{GSL IN $f(G)$-GRAVITY}

Here, we are interested in exploring the GSL of gravitational
thermodynamics in the context of $f(G)$-gravity. To this aim, we
consider a spatially flat FRW universe filled with the matter and
radiation. We further assume that the boundary of the universe to be
enclosed by the dynamical apparent horizon with the Hawking
temperature. For a spatially flat FRW universe, the dynamical
apparent horizon takes the form \cite{Poisson,Cai09}
\begin{equation}\label{ra}
\tilde{r}_{\rm A}=H^{-1},
\end{equation}
which is same as the Hubble horizon. Following \cite{Cai05}, the
Hawking temperature on $\tilde{r}_{\rm A}$ is given by

\begin{equation}
T_{\rm A}=\frac{1}{2\pi \tilde{r}_{\rm A}}
\left(1-\frac{\dot{\tilde{r}}_{\rm A}}{2H\tilde{r}_{\rm
A}}\right).\label{ta}
\end{equation}
Now we are going to use the first law of thermodynamics to find the
general condition needed to hold the GSL in $f(G)$-gravity. The
entropy of matter and radiation inside the horizon are given by the
Gibbs equation \cite{Izquierdo1}

\begin{equation}\label{TdSm}
T_{\rm A}{\rm d}S_{\rm m}={\rm d}E_{\rm m}+p_{\rm m}{\rm d}{V},
\end{equation}

\begin{equation}\label{TdSr}
T_{\rm A}{\rm d}S_{\rm r}={\rm d}E_{\rm r}+p_{\rm r}{\rm d}{V},
\end{equation}
where $E_{\rm m}=\rho_{\rm m}{V}$ and $E_{\rm r}=\rho_{\rm r}{V}$.
Also $V= 4\pi\tilde{r}_{\rm A}^{3}/3$ is the volume of the dynamical
apparent horizon $\tilde{r}_{\rm A}$ containing the pressureless
matter ($p_{\rm m}=0$) and radiation ($p_{\rm r}=\rho_{\rm r}/3$).

Taking time derivative of Eqs. (\ref{TdSm}) and (\ref{TdSr}) and
using (\ref{co eq2}) and (\ref{co eq1}) one can get
\begin{equation}\label{S mat f}
T_{\rm A}\dot{S}=4\pi\tilde{r}_{\rm A}^{2}\left(\rho_{\rm m}
+\frac{4}{3}\rho_{\rm r}\right)\left(\dot{\tilde{r}_{\rm
A}}-H\tilde{r}_{\rm A}\right),
\end{equation}
where $S=S_{\rm r}+S_{\rm m}$. Replacing $\rho_{\rm m}$ and
$\rho_{\rm r}$ from Eqs. (\ref{fr 1}) and (\ref{fr 2}) into the
above relation and using $\tilde{r}_{\rm A}=H^{-1}$, one can obtain
\begin{equation} T_{\rm
A}\dot{S}=\frac{2\pi\left(\dot{H}+H^{2}\right)}{H^{4}}
\Big[4\dot{H}-16H\left(H^{2}-2\dot{H}\right)\dot{f}_{\rm
G}+16H^{2}\ddot{f}_{\rm G} \Big].\label{sma}
\end{equation}
The horizon entropy in $f(R,G)$-gravity is given by \cite{Wald}
\begin{equation}\label{S hor}
S_{\rm A}= -\pi \int\left(F_{\rm R}\frac{\partial R}{\partial
R_{\alpha\beta\gamma\rho}} + F_{\rm G}\frac{\partial G}{\partial
R_{\alpha\beta\gamma\rho}}\right)\varepsilon_{\alpha\beta}\varepsilon_{\gamma\rho}{\rm
d}A,
\end{equation}
where $F_{\rm R}=\frac{\partial F(R,G)}{\partial R}$, $F_{\rm
G}=\frac{\partial F(R,G)}{\partial G}$ and ${\rm
A}=4\pi\tilde{r}_{\rm A}^{2}$ is the area of the apparent horizon
($\tilde{r}_{\rm A}=H^{-1}$). Also the quantity
$\varepsilon_{\alpha\beta}$ is normalized as
$\varepsilon^{\alpha\beta}\varepsilon_{\alpha\beta}=-2$ and
antisymmetric under the exchange $\alpha\longleftrightarrow\beta$.
It is the binormal vector to the bifurcation surface \cite{Wald}.
For the action (\ref{action1}) we have
\begin{equation}
F(R,G)= R + 2f(G),\label{fR}
\end{equation}
therefore Eq. (\ref{S hor}) yields
\begin{equation}\label{S hor1}
S_{\rm A}= 8\pi^{2} \left(H^{-2}+ 8f_{\rm G}\right).
\end{equation}
Taking time derivative of Eq. (\ref{S hor1}) and using the Hawking
temperature (\ref{ta}), one can derive the evolution of the horizon
entropy as
\begin{equation}\label{saa}
T_{\rm A}\dot{S}_{\rm
A}=\frac{2\pi\left(\dot{H}+2H^{2}\right)}{H^{4}}\left(-2\dot{H}
+8H^{3}\dot{f}_{\rm G}\right).
\end{equation}
Summing up Eqs. (\ref{sma}) and (\ref{saa}), the GSL in
$f(G)$-gravity yields
\begin{equation}\label{S Tot}
T_{\rm A}\dot{S}_{\rm tot}= \frac{2\pi}{H^{4}}\left[2\dot{H}^{2}
+8H\dot{H}\Big(4\dot{H}+3H^{2}\Big)\dot{f}_{\rm G}
+16H^{2}\Big(\dot{H}+H^{2}\Big)\ddot{f}_{\rm G}\right],
\end{equation}
where $S_{\rm tot}=S_{\rm r}+S_{\rm m}+S_{\rm A}$. Equation (\ref{S
Tot}) shows that in $f(G)$ gravity, the validity of the GSL, i.e.
$T_{\rm A}\dot{S}_{\rm tot}\geq 0$, depends on the explicit form of
the $f(G)$ model. For the Einstein gravity ($f(G)=0$), one can
immediately find that the GSL (\ref{S Tot}) reduces to
\begin{equation}\label{GSL-R}
T_{\rm A}\dot{S}_{\rm tot}=\frac{4\pi\dot{H}^{2}}{H^{4}}\geq 0,
\end{equation}
which shows that the GSL is always fulfilled throughout history of
the universe. In section \ref{two viable}, we examine the validity
of the GSL (\ref{S Tot}) for two viable $f(G)$-gravity models.

\section{Dynamics of $f(G)$-gravity}\label{dynamics}

To study the dynamics of a general $f(G)$ model, we use the
following dimensionless variables \cite{JCAP,Halliwell}

\begin{equation}\label{X1}
x_{1}=\frac{Gf_{\rm
G}}{3H^{2}},~~~~~~~~~~~~~~~~~~~~~~~~~~~~~~~~~~~~~~~~~
\end{equation}

\begin{equation}\label{X2}
x_{2}=-\frac{f}{3H^{2}},~~~~~~~~~~~~~~~~~~~~~~~~~~~~~~~~~~~~~~~
\end{equation}

\begin{equation}\label{X3}
x_{3}=-8H\dot{f}_{\rm G},~~~~~~~~~~~~~~~~~~~~~~~~~~~~~~~~~~~~~~
\end{equation}

\begin{equation}\label{X4}
x_{4}=\Omega_{\rm r}=\frac{\rho_{\rm
r}}{3H^{2}},~~~~~~~~~~~~~~~~~~~~~~~~~~~~~~~~~~~
\end{equation}

\begin{equation}\label{X5}
x_{5}=\frac{G}{24H^{4}}=\frac{\dot{H}}{H^{2}}+1,~~~~~~~~~~~~~~~~~~~~~~~~~~
\end{equation}

\begin{equation}\label{X6}
x_6=H.~~~~~~~~~~~~~~~~~~~~~~~~~~~~~~~~~~~~~~~~~~~~~
\end{equation}
With the help of above definitions and using Eqs. (\ref{fr 1}),
(\ref{fr 2}), (\ref{co eq2}) and (\ref{co eq1}) one can get a set of
first order differential equations governing a general $f(G)$ model
as \cite{JCAP}

\begin{equation}\label{dx 1}
\frac{{\rm d}x_{1}}{{\rm d}N}=
-x_{3}x_{5}\left(1+\frac{1}{m}\right)+2x_{1}(1-x_5),~~~~~~~~~~~~~~~~~~~~~
\end{equation}

\begin{equation} \label{dx 2}
\frac{{\rm d}x_{2}}{{\rm d}N}=
\frac{x_{3}x_{5}}{m}+2x_{2}(1-x_5),~~~~~~~~~~~~~~~~~~~~~~~~~~~~~~~~~~~
\end{equation}

\begin{equation}\label{dx 3}
\frac{{\rm d}x_{3}}{{\rm d}N}=
-x_{3}(1+x_5)+2x_{5}+1-3(x_{1}+x_{2})+x_{4},~~~~~~~
\end{equation}

\begin{equation} \label{dx 4}
\frac{{\rm d}x_{4}}{{\rm d}N}=
-2x_{4}(1+x_{5}),~~~~~~~~~~~~~~~~~~~~~~~~~~~~~~~~~~~~~~~~~~
\end{equation}

\begin{equation} \label{dx 5}
\frac{{\rm d}x_{5}}{{\rm d}N}=
x_5\left[4(1-x_5)-\frac{x_{3}x_{5}}{mx_{1}}\right],~~~~~~~~~~~~~~~~~~~~~~~~~~~~~~
\end{equation}

\begin{equation} \label{dx 6}
\frac{{\rm d}x_{6}}{{\rm d}N}=
(x_5-1)x_6,~~~~~~~~~~~~~~~~~~~~~~~~~~~~~~~~~~~~~~~~~~~~~~
\end{equation}
where $N=\ln(a/a_{i})$ and $a_{i}$ is the initial value of the
scalar factor. Also
\begin{equation}\label{m}
m=\frac{Gf_{\rm GG}}{f_{\rm G}}.~~~~~~~~~~~~~~~~~~~~~~~
\end{equation}
Notice that the variable $x_6=H$, Eq. (\ref{X6}), and the
differential equation (\ref{dx 6}) are absent in \cite{JCAP}.

Using Eqs. (\ref{X1})-(\ref{X4}), one can rewrite Eqs. (\ref{fr 1})
and (\ref{ro G}) as
\begin{equation}\label{Omega m} \Omega_{\rm
m}=1-x_{1}-x_{2}-x_{3}-x_{4},
\end{equation}
\begin{equation}\label{Omega G}
\Omega_{\rm G}=x_{1}+x_{2}+x_{3},
\end{equation}
where $\Omega_{\rm m}=\frac{\rho_{\rm m}}{3H^{2}}$ and $\Omega_{\rm
G}=\frac{\rho_{\rm G}}{3H^{2}}$. Also we have $\Omega_{\rm
m}+\Omega_{\rm r}+\Omega_{\rm G}=1$.

From Eqs. (\ref{w DE}) and (\ref{w eff}), one can rewrite
$\omega_{\rm G}$ and $\omega_{\rm eff}$ in terms of the variables
$x_i$ as

\begin{equation}\label{omega G}
\omega_{\rm G} =\frac{-2x_{5}-x_{4}-1}{3(x_{1}+x_{2}+x_{3})},
\end{equation}

\begin{equation}\label{omega eff}
\omega_{\rm eff}=-\frac{1}{3}(2x_{5}+1).
\end{equation}
Also the deceleration parameter takes the form
\begin{equation}\label{q}
q=-1-\frac{\dot{H}}{H^{2}}=-x_{5},
\end{equation}
and
\begin{equation}\label{H^6 fgg}
H^{6}f_{\rm GG}=\frac{mx_{1}}{192~x_{5}^{2}}.
\end{equation}
Notice that in the context of $f(G)$-gravity, the quantity
$H^{6}f_{\rm GG}$ plays an important role. In a viable $f(G)$ model,
the condition $0 < H^6 f_{\rm GG} < 1/384$ is necessary in order to
have a stable de Sitter point \cite{PLB}.

From Eqs. (\ref{ro G}) and (\ref{p G}), the energy density and
pressure due to the $f(G)$ contribution can be expressed in terms of
the variables $x_i$ as

\begin{equation}\label{ro G1}
\rho_{\rm G} = 3H^{2}\left(x_{1}+x_{2}+x_{3}\right),
\end{equation}

\begin{equation}\label{p G1}
p_{\rm G}=-H^{2}\left(2x_{5}+x_{4}+1\right).
\end{equation}
With the help of Eqs. (\ref{X1})-(\ref{dx 6}), the GSL (\ref{S Tot})
reads
\begin{equation}\label{GSL 2}
T_{\rm A}\dot{S}_{\rm tot} =2\pi\big\{ 2x_{5} \big[
3(x_{1}+x_{2}-1)-(x_{4}+x_{5}) \big] +x_{3}(5x_{5}-1)+2 \big\}.
\end{equation}
Therefore for a given $f(G)$ model, solving the set of first order
differential equations (\ref{dx 1})-(\ref{dx 6}) numerically, one
can obtain the evolutionary behaviours of $H$, $G$, $\Omega_{\rm
m}$, $\Omega_{\rm G}$, $\omega_{\rm G}$, $\omega_{\rm eff}$, $q$ and
$T_{\rm A}\dot{S}_{\rm tot}$. In what follows, we investigate the
dynamics of two viable $f(G)$ models and examine the validity of the
GSL, i.e. $T_{\rm A}\dot{S}_{\rm tot}\geq 0$.

\section{Two viable $f(G)$ models}\label{two viable}

Here, we are interested in investigating the GSL in two viable
$f(G)$ models. The first model has the form \cite{JCAP}
\begin{equation} \label{f(g)A}
f(G)=\alpha\left(G^{\frac{3}{4}}
-\beta\right)^{\frac{2}{3}},~~~~~~{\rm Model~I},
\end{equation}
where $\alpha$ and $\beta$ are two constants of the model. The
second $f(G)$ model is given by \cite{PLB}

\begin{equation}\label{f(g)B}
f(G)=\lambda\frac{G}{\sqrt{G_{\ast}}}\arctan\left(\frac{G}{G_{\ast}}\right)
-\frac{\lambda}{2}\sqrt{G_{\ast}}~\ln\left(1+\frac{G^{2}}{G^{2}_{\ast}}\right)-\alpha\lambda\sqrt{G_{\ast}},~~~~~~{\rm
Model~II},
\end{equation}
where $\alpha$ is an arbitrary constant and $\lambda$ is a positive
constant. Also $G_{\ast}=H^{4}_{0}$ and $H_0 $ is the Hubble
parameter at present.

\begin{figure}
\includegraphics{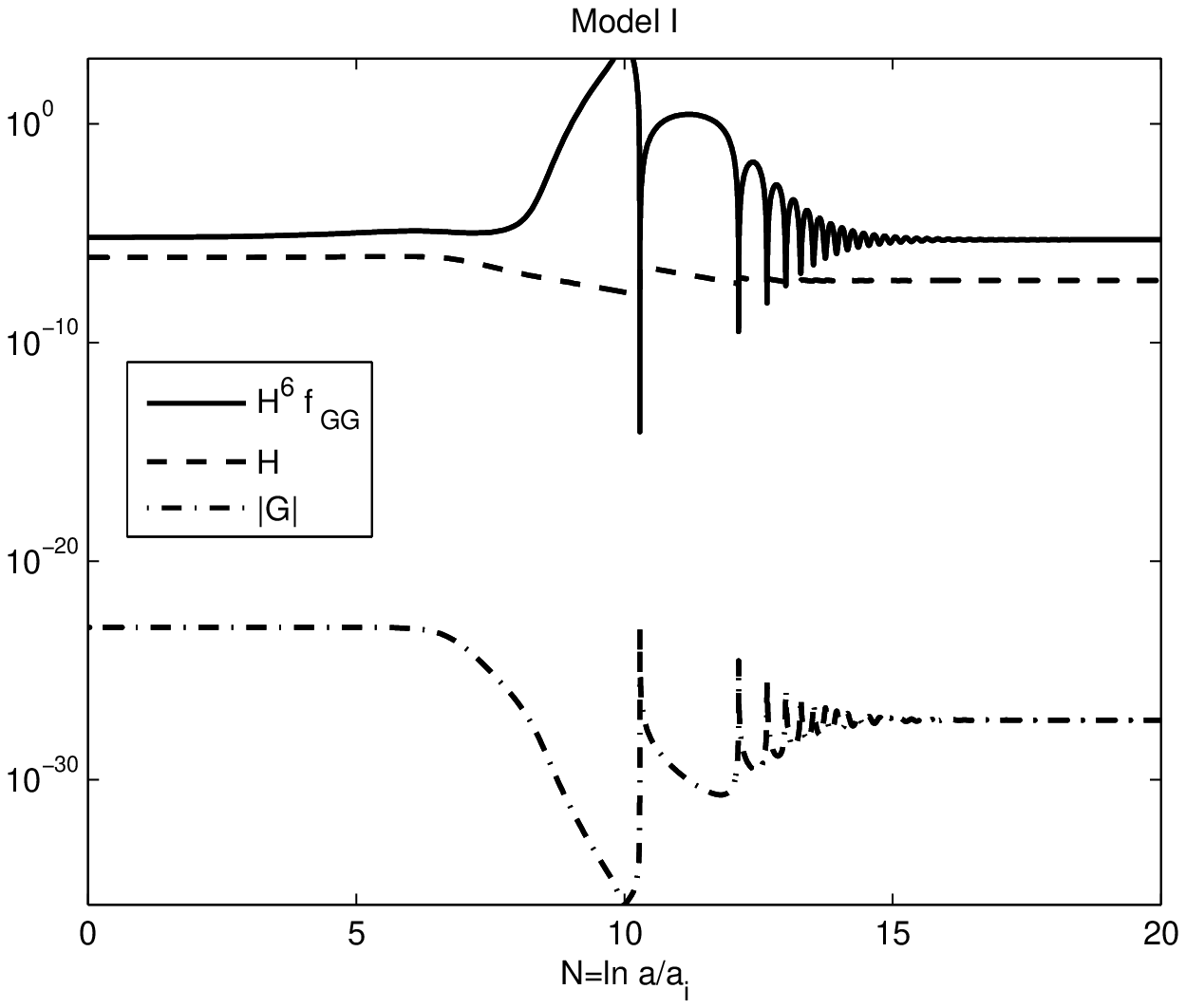} \vspace{4.0cm} \caption[]{The evolutions of the Hubble
parameter $H$, the Gauss-Bonnet curvature invariant term $|G|$ and
the quantity $H^{6}f_{\rm GG}$, Eqs. (\ref{X5}), (\ref{X6}) and
(\ref{H^6 fgg}), versus $N=\ln(a/a_{\rm i})$ where $a_{\rm i}$ is
the initial value of the scale factor. Auxiliary parameters are:
$\alpha=\frac{1}{40\sqrt{66}}$ and
      $\beta = -10^{-17}$. Initial values are: $x_{1}=-0.0025$,
      $x_{2}=0.005$, $x_{3}=-0.01$, $x_{4}=0.99951$, $x_{5}=-0.99$ \cite{JCAP} and
      $x_{6}=7.95225\times10^{-7}$. }
\label{JCAP2}
\end{figure}

\begin{figure}
\includegraphics{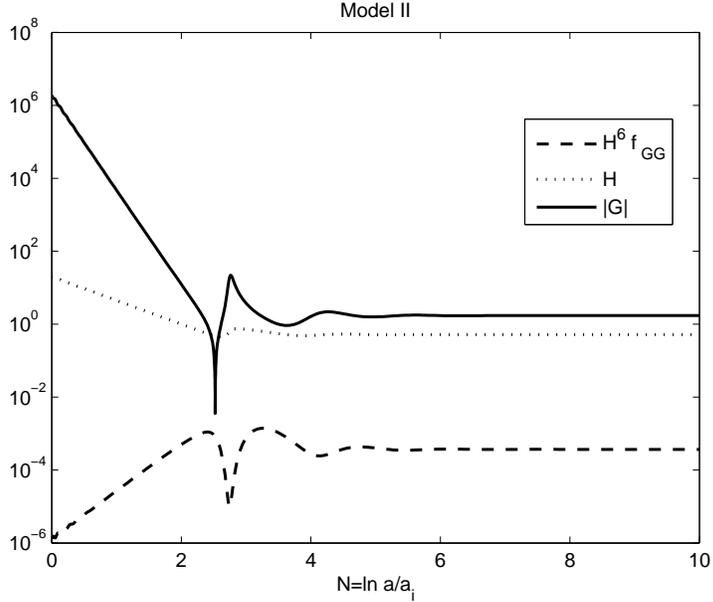} \vspace{4.0cm} \caption[]{Same as Fig. \ref{JCAP2} but
for model II. Auxiliary parameters are: $\alpha=10$ and $\lambda =
0.075$ \cite{PLB}. Initial
      values are: $x_{1}=189.249 $, $x_{2}=-189.248$, $x_{3}=-0.0014$, $x_{4}=0.004$, $x_{5}=-0.502$
      and $x_{6}=20$.}
\label{PLB2}
\end{figure}

With choice of suitable initial conditions, we numerically solve the
differential equations (\ref{dx 1})-(\ref{dx 6}) for both model I
and model II. The evolutions of the Hubble parameter $H$, the
Gauss-Bonnet curvature invariant term $|G|$ and the quantity
$H^6f_{\rm GG}$, Eqs. (\ref{X5}), (\ref{X6}) and (\ref{H^6 fgg}),
versus $N=\ln(a/a_{\rm i})$ for model I and model II are plotted in
Figs. \ref{JCAP2} and \ref{PLB2}, respectively. Figures show that:
(i) the Hubble parameter decreases during history of the universe.
(ii) The Gauss-Bonnet curvature invariant term changes its sign when
it transits from the standard radiation/matter dominated epochs to
the accelerated era. (iii) The quantity $H^6f_{\rm GG}$ satisfies
the condition $0 < H_1^6 f_{\rm GG}(G_1) < 1/384$ which shows that
both models have a stable de Sitter attractor. (iv) $H$, $|G|$ and
$H^6f_{\rm GG}$ at late times go to a constant value when the
universe enters a de Sitter regime. Notice that the result of Fig.
\ref{PLB2} for model II is the same as that obtained in \cite{PLB}.

\begin{figure}
\includegraphics{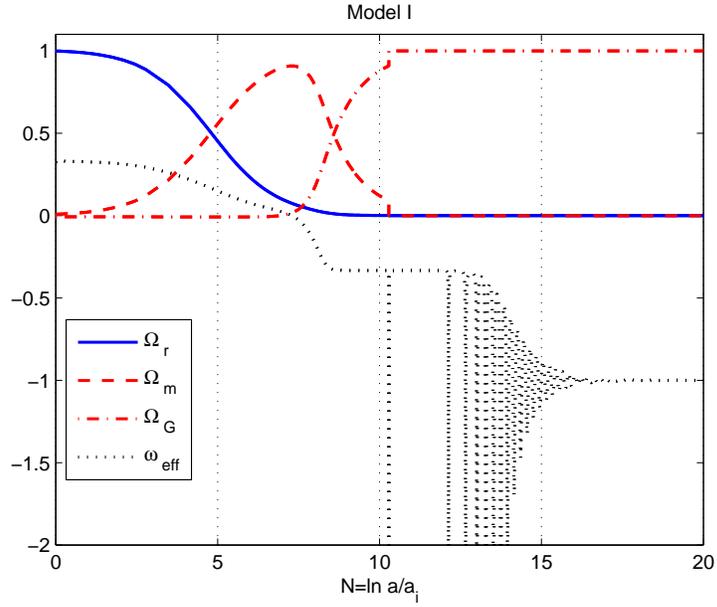} \vspace{4.0cm} \caption[]{The evolutions of
$\Omega_{\rm m}$, $\Omega_{\rm G}$, $\Omega_{\rm r}$ and
$\omega_{\rm eff}$, Eqs. (\ref{Omega m}), (\ref{Omega G}) and
(\ref{omega eff}), versus $N$.  Auxiliary parameters and initial
values as in Fig. \ref{JCAP2}.} \label{JCAP3}
\end{figure}

\begin{figure}
\includegraphics{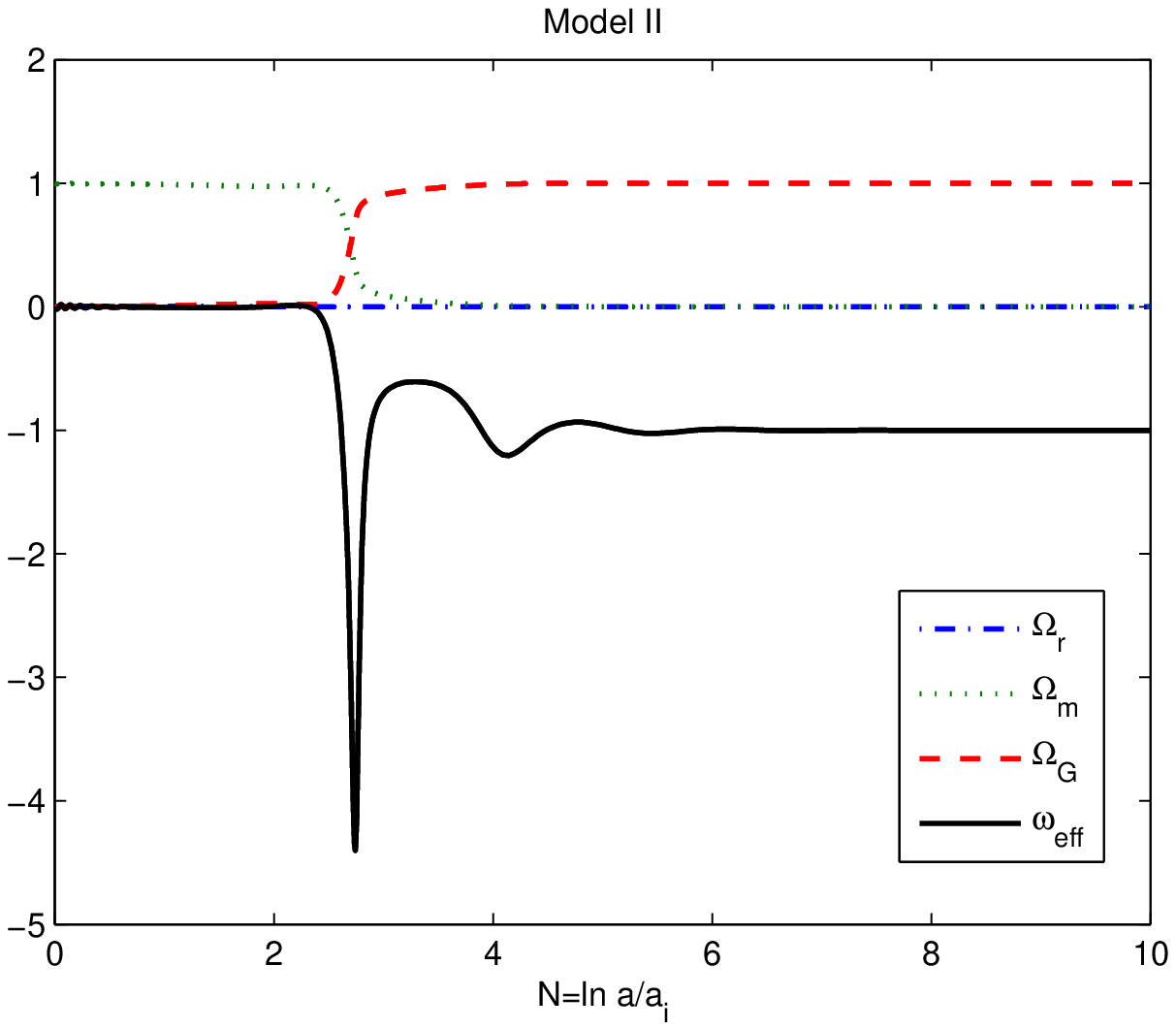} \vspace{4.0cm} \caption[]{Same as Fig. \ref{JCAP3} but
for model II. Auxiliary parameters and initial values as in Fig.
\ref{PLB2}.} \label{PLB3}
\end{figure}

The evolutions of the density parameters $\Omega_{\rm r}$,
$\Omega_{\rm m}$, $\Omega_{\rm G}$ and the effective EoS parameter
$\omega_{\rm eff}$, Eqs. (\ref{Omega m}), (\ref{Omega G}) and
(\ref{omega eff}), versus $N$ for model I and model II are plotted
in Figs. \ref{JCAP3} and \ref{PLB3}, respectively. Figures
illustrate that: (i) for both models, $\Omega_{\rm r}$, $\Omega_{\rm
m}$, $\Omega_{\rm G}$ and $\omega_{\rm eff}$ behave like the
$\Lambda$CDM model in the radiation/matter dominated epochs. (ii)
For model I, $\omega_{\rm eff}$ oscillates rapidly during the
accelerated epoch and goes deep into the phantom-like region as the
universe enters the de Sitter period. (iii) For model II,
$\omega_{\rm eff}$ oscillates slowly around $-1$ as the system
enters the epoch of cosmic acceleration, which implies that the de
Sitter solution is a stable spiral. Note that the results of Figs.
\ref{JCAP3} and \ref{PLB3} are the same as those obtained in
\cite{JCAP} and \cite{PLB}, respectively.

\begin{figure}
\includegraphics{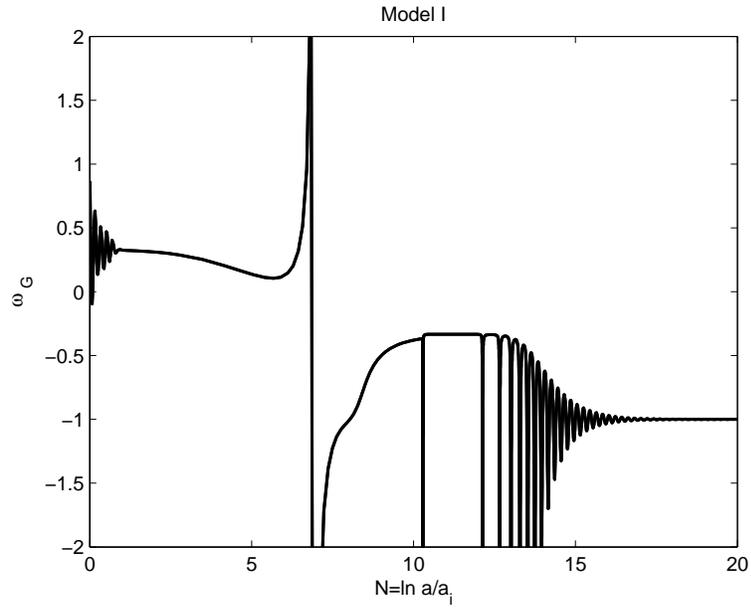}
      \vspace{4.0cm}
      \caption[]{The evolution of the EoS parameter $\omega_{\rm G}$, Eq. (\ref{omega G}), versus $N$ for model I.
       Auxiliary parameters and initial
values as in Fig. \ref{JCAP2}.}
         \label{JCAP4}
   \end{figure}

\begin{figure}
\includegraphics{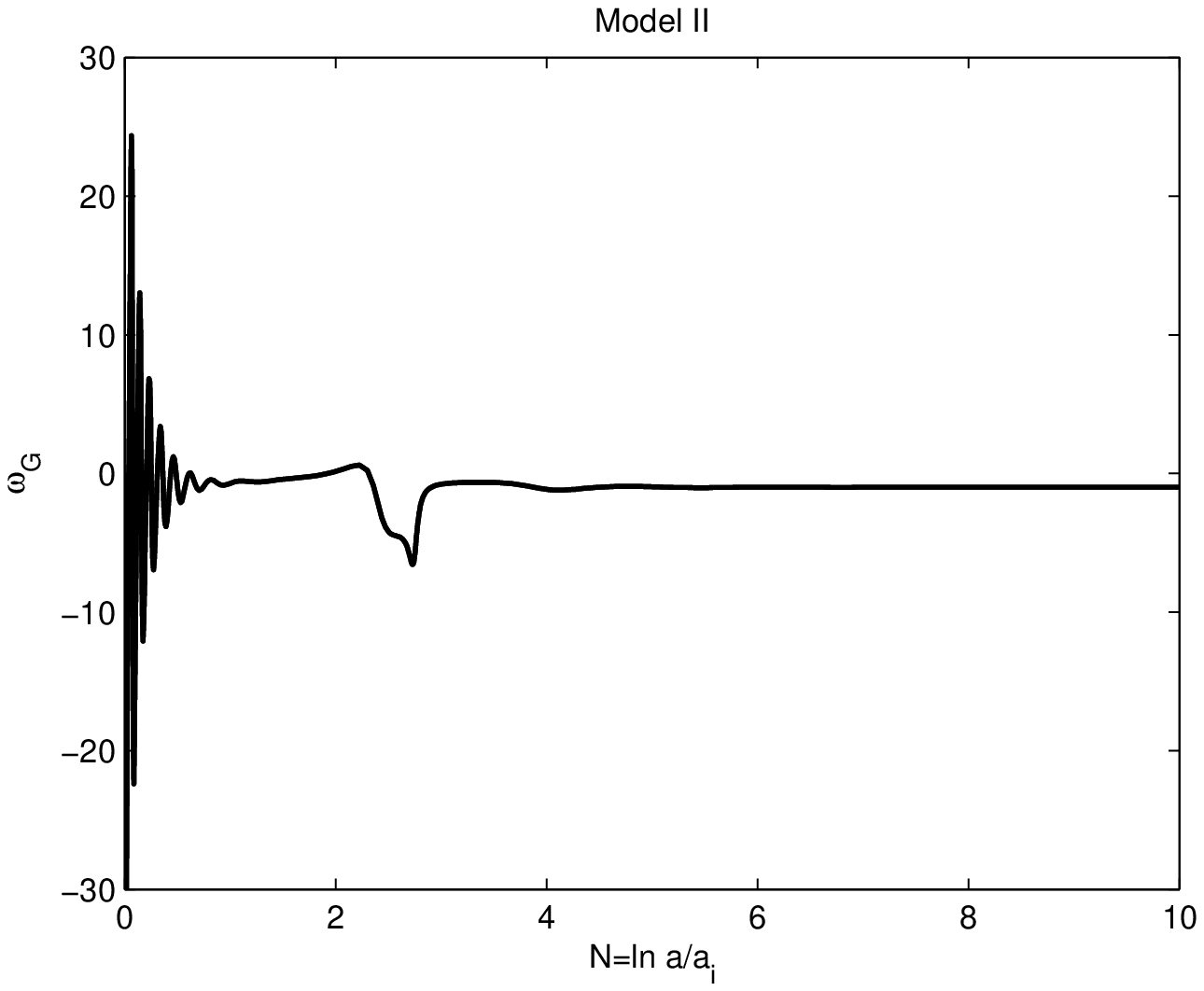}
      \vspace{4.0cm}
      \caption[]{Same as Fig. \ref{JCAP4} but for model II. Auxiliary parameters and initial values as in Fig.
\ref{PLB2}.}
         \label{PLB4}
   \end{figure}

The evolutionary behavior of the EoS parameter $\omega_{\rm G}$ due
to the $f(G)$ contribution, Eq. (\ref{omega G}), for model I and
model II is plotted in Figs. \ref{JCAP4} and \ref{PLB4},
respectively. Figures present that $\omega_{\rm G}$ oscillates
quickly at early times and approaches a de Sitter regime at late
times, as expected.

\begin{figure}
\includegraphics{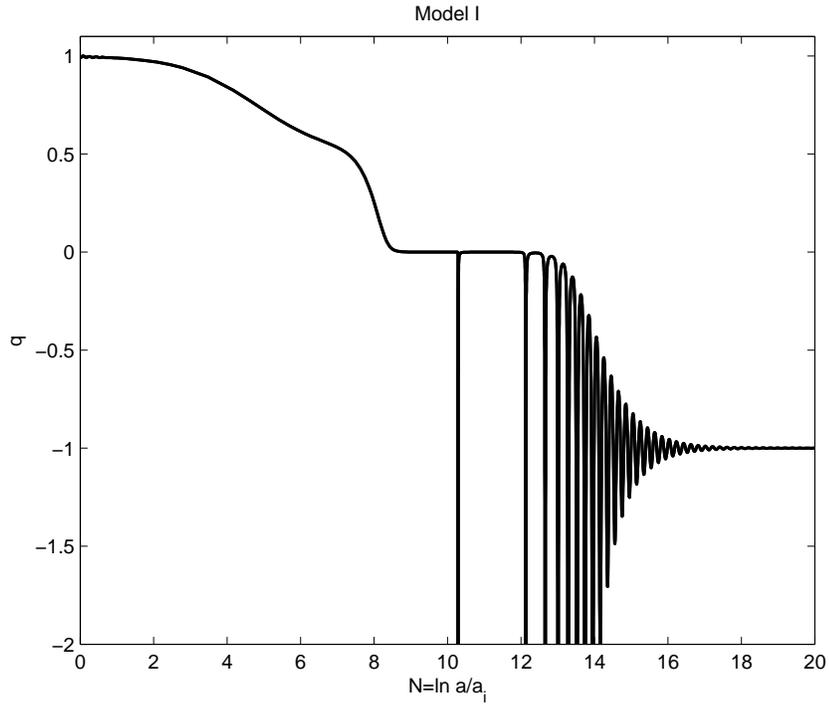}
      \vspace{5cm}
      \caption[]{The evolution of the deceleration parameter $q$, Eq. (\ref{q}), versus $N$ for model I.
      Auxiliary parameters and initial values as in Fig. \ref{JCAP2}.}
         \label{JCAP5}
   \end{figure}

\begin{figure}
\includegraphics{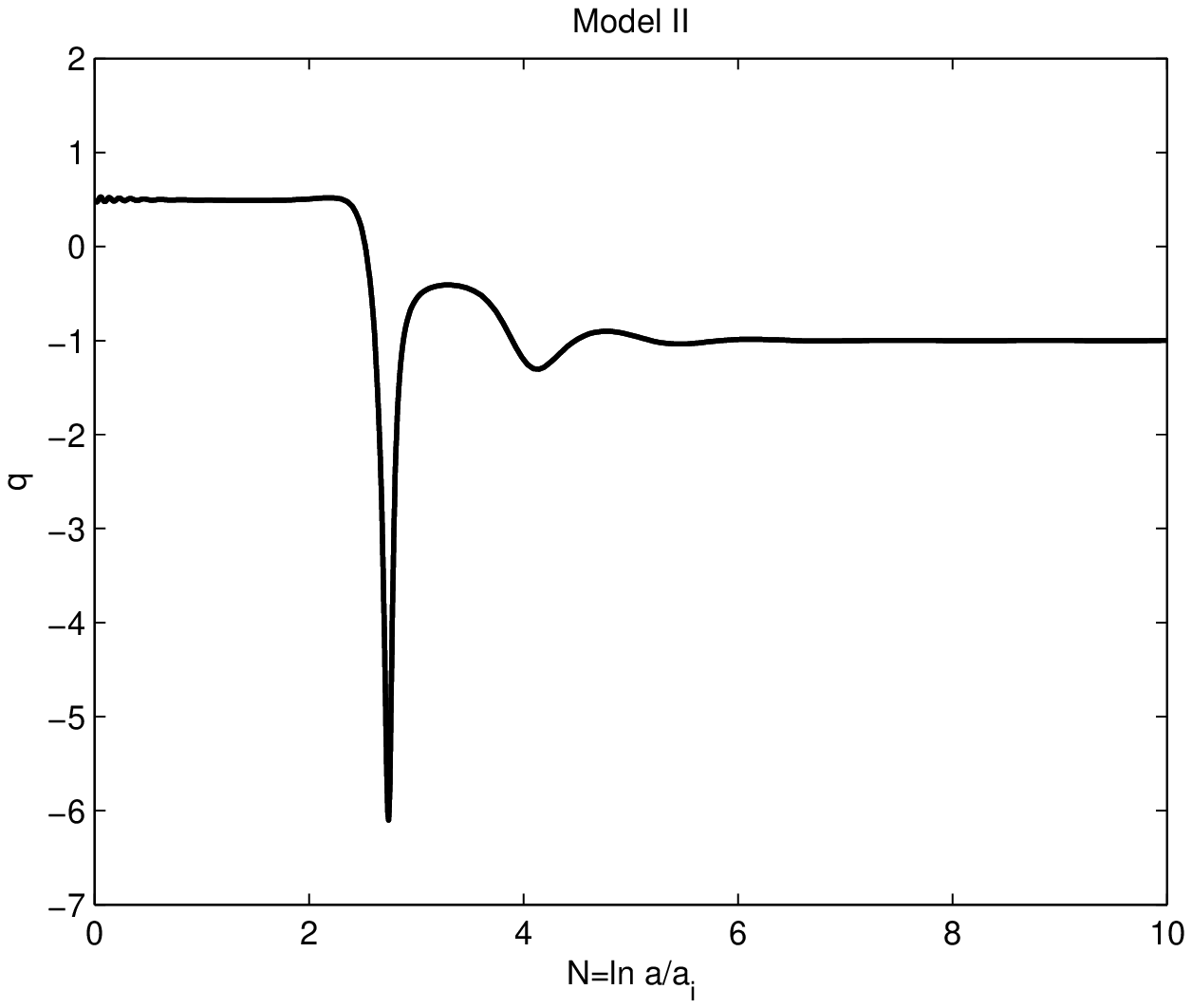}
      \vspace{4.0cm}
      \caption[]{Same as Fig. \ref{JCAP5} but for model II. Auxiliary parameters and initial values as in Fig. \ref{PLB2}.}
         \label{PLB5}
   \end{figure}

The evolution of the deceleration parameter $q$, Eq. (\ref{q}), for
model I and model II is plotted in Figs. \ref{JCAP5} and \ref{PLB5},
respectively. Figure \ref{JCAP5} clears that for model I, the
deceleration parameter starts from $q=1$ corresponding to the
radiation dominated epoch, then shows a cosmic deceleration ($q>0$)
to acceleration ($q<0$) transition \cite{Ishida} and finally
oscillates rapidly into the de Sitter regime ($q=-1$). Figure
\ref{PLB5} presents that for model II, $q$ varies from the matter
dominated epoch ($q=0.5$), then transits from a cosmic deceleration
to acceleration and approaches smoothly a de Sitter regime at late
times, as expected.

In addition, we turn to check the energy conditions in both model I
and model II. The energy conditions are as follows
\cite{HE,Bamba,Barrow}:

\noindent~(i) The null energy condition (NEC), i.e. $\rho_{\rm
G}+p_{\rm G}\geq 0$.

\noindent~(ii) The strong energy condition (SEC), i.e. $\rho_{\rm
G}+p_{\rm G}\geq 0$ and $\rho_{\rm G}+3p_{\rm G}\geq 0$.

\noindent~(iii) The weak energy condition (WEC), i.e. $\rho_{\rm
G}+p_{\rm G}\geq 0$ and $\rho_{\rm G}\geq 0$.

\noindent~(iv) The dominant energy condition (DEC), i.e. $\rho_{\rm
G}\geq 0$ and $\rho_{\rm G}\geq |p_{\rm G}|$.

\begin{figure}
\includegraphics{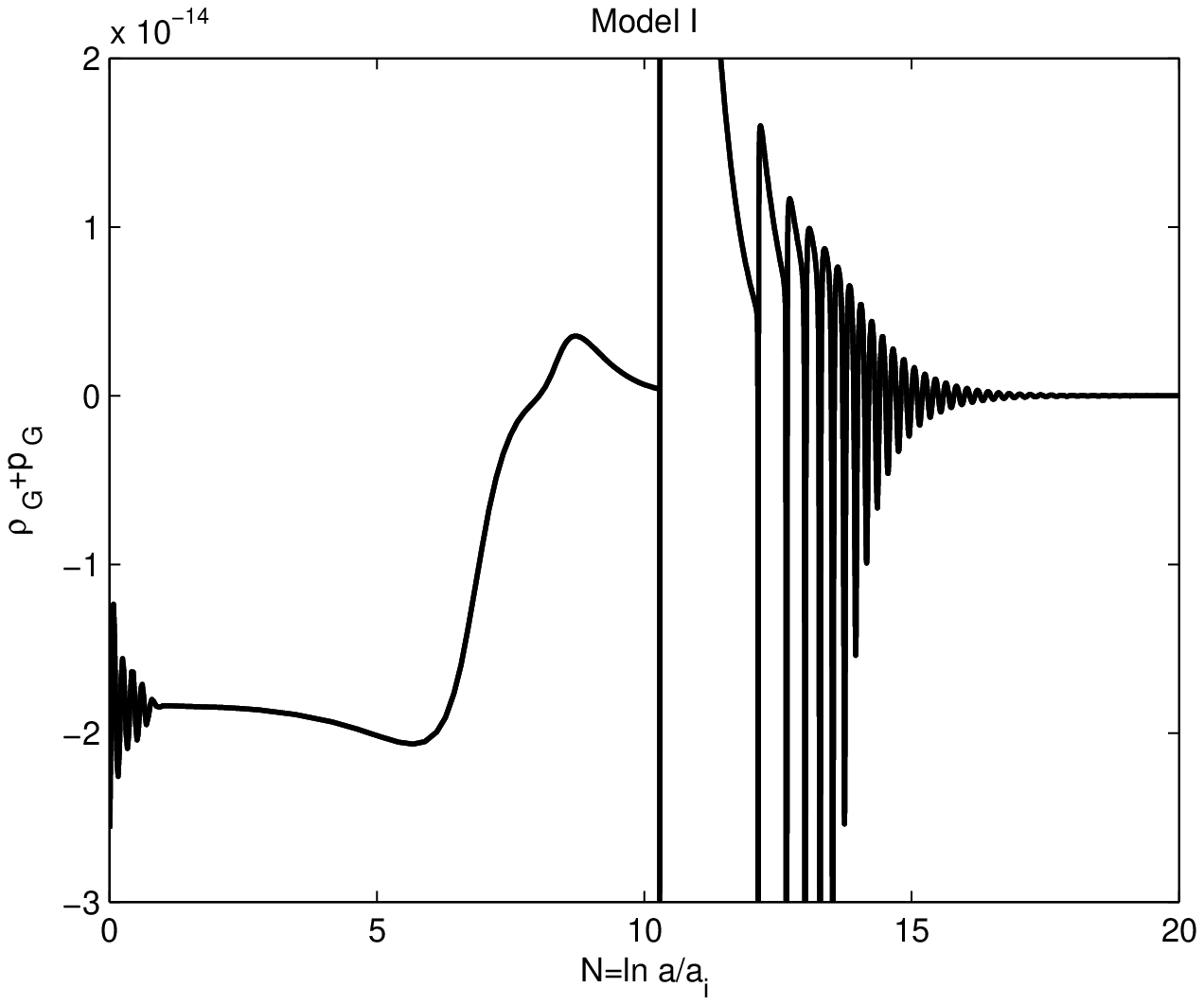}
      \vspace{4.0cm}
      \caption[]{The evolution of $\rho_{\rm G}+p_{\rm G}$, Eqs. (\ref{ro G1}) and (\ref{p G1}),
      versus $N$ for model I. Auxiliary parameters and initial values as in Fig. \ref{JCAP2}.}
         \label{JCAP7}
   \end{figure}

\begin{figure}
\includegraphics{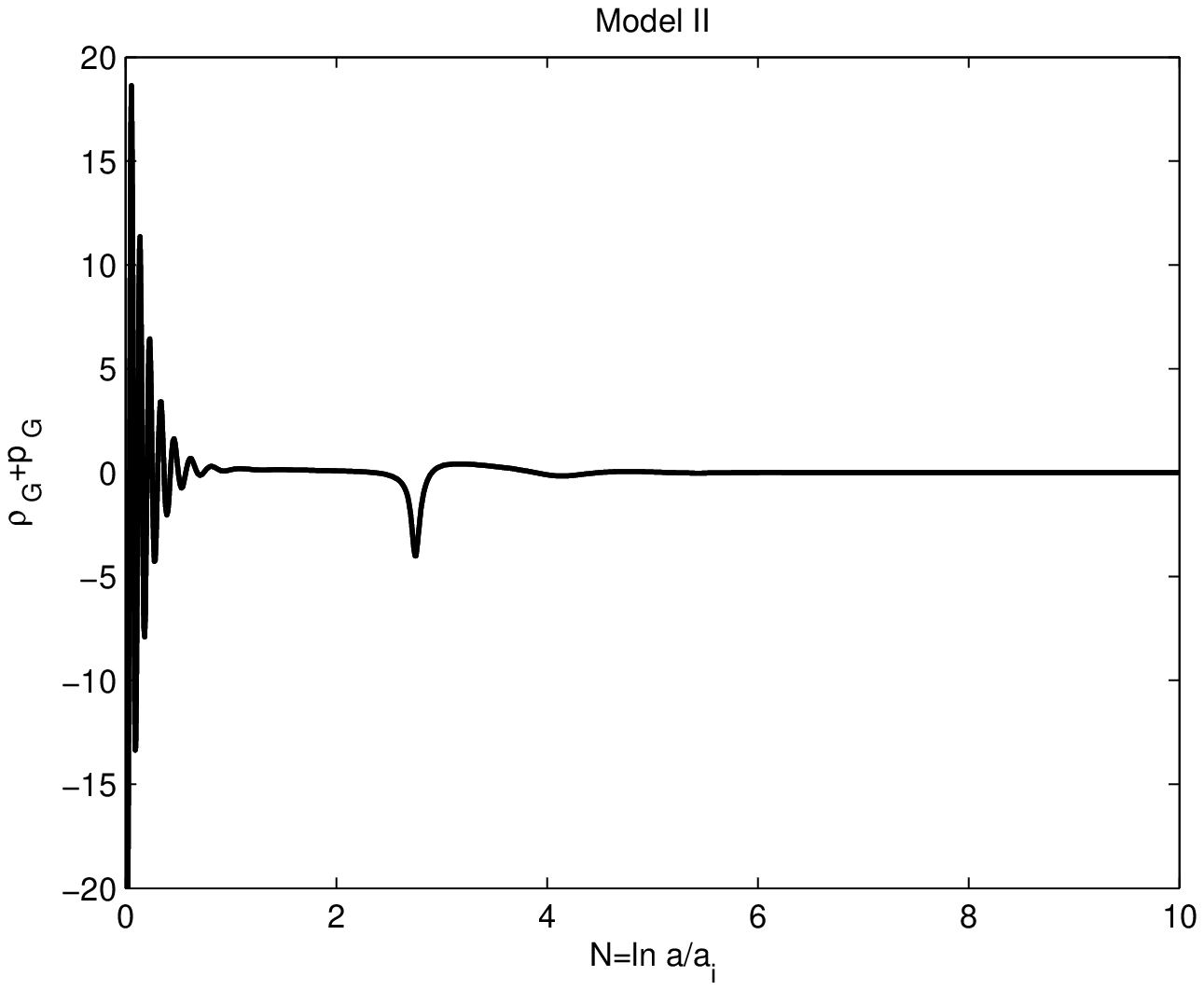}
      \vspace{4.0cm}
      \caption[]{Same as Fig. \ref{JCAP7} but for model II. Auxiliary parameters and initial values as in Fig. \ref{PLB2}.}
         \label{PLB7}
   \end{figure}

\begin{figure}
\includegraphics{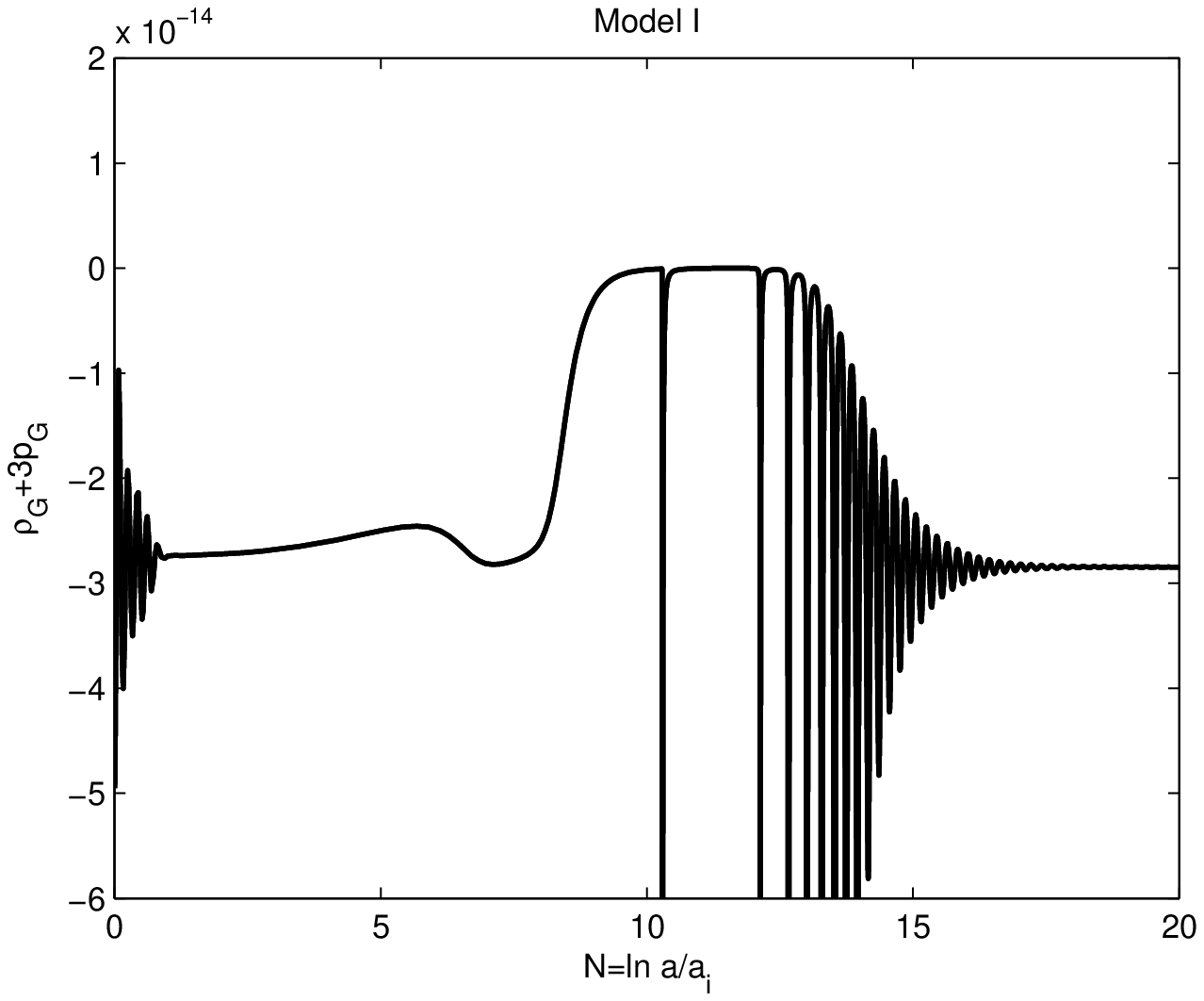}
      \vspace{4.0cm}
      \caption[]{The evolution of $\rho_{\rm G}+3p_{\rm G}$, Eqs. (\ref{ro G1}) and (\ref{p G1}), versus $N$ for model I.
      Auxiliary parameters and initial values as in Fig. \ref{JCAP2}.}
         \label{JCAP8}
   \end{figure}

\begin{figure}
\includegraphics{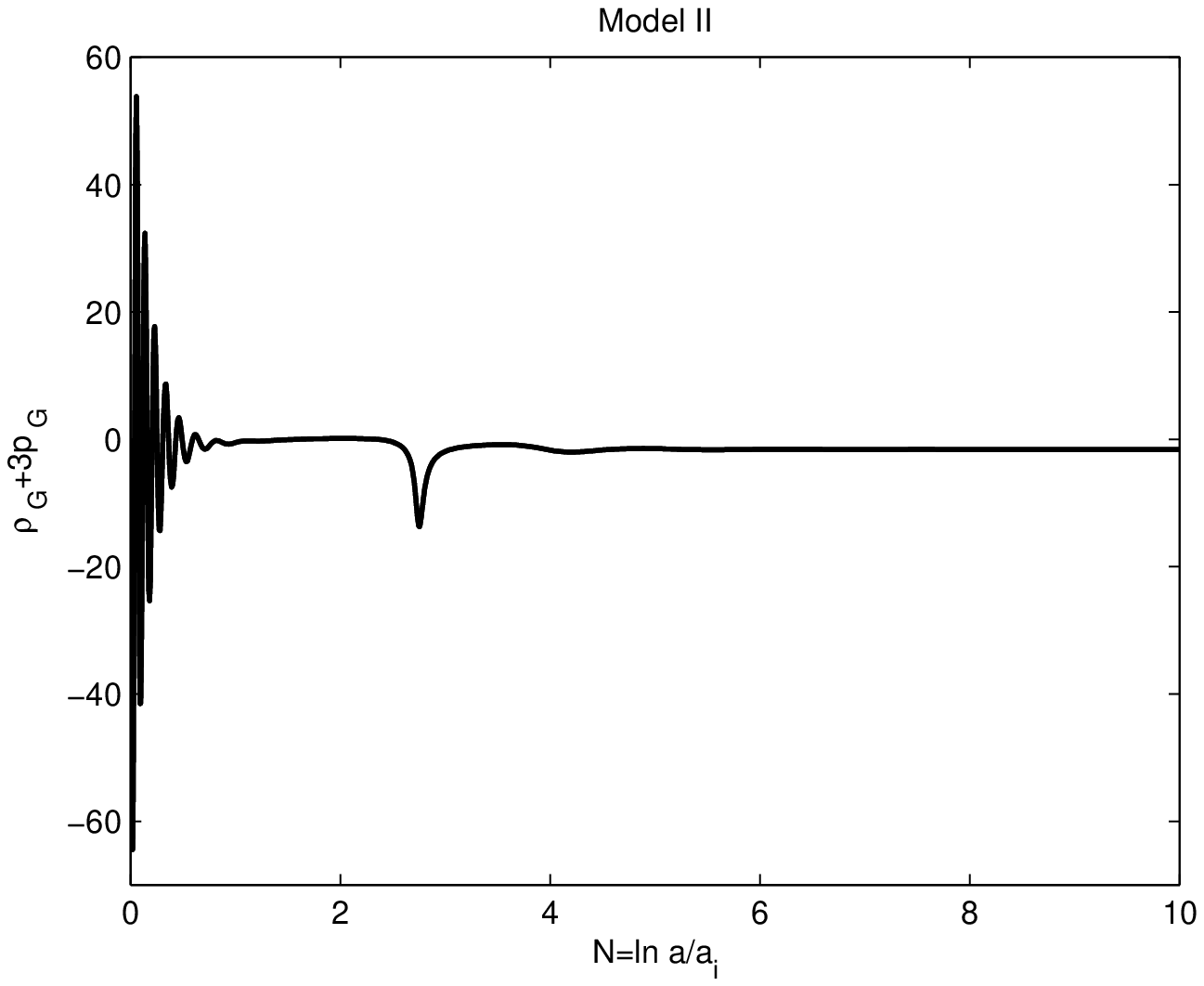}
      \vspace{4.0cm}
      \caption[]{Same as Fig. \ref{JCAP8} but for model II. Auxiliary parameters and initial values as in Fig. \ref{PLB2}.}
         \label{PLB8}
   \end{figure}

\begin{figure}
\includegraphics{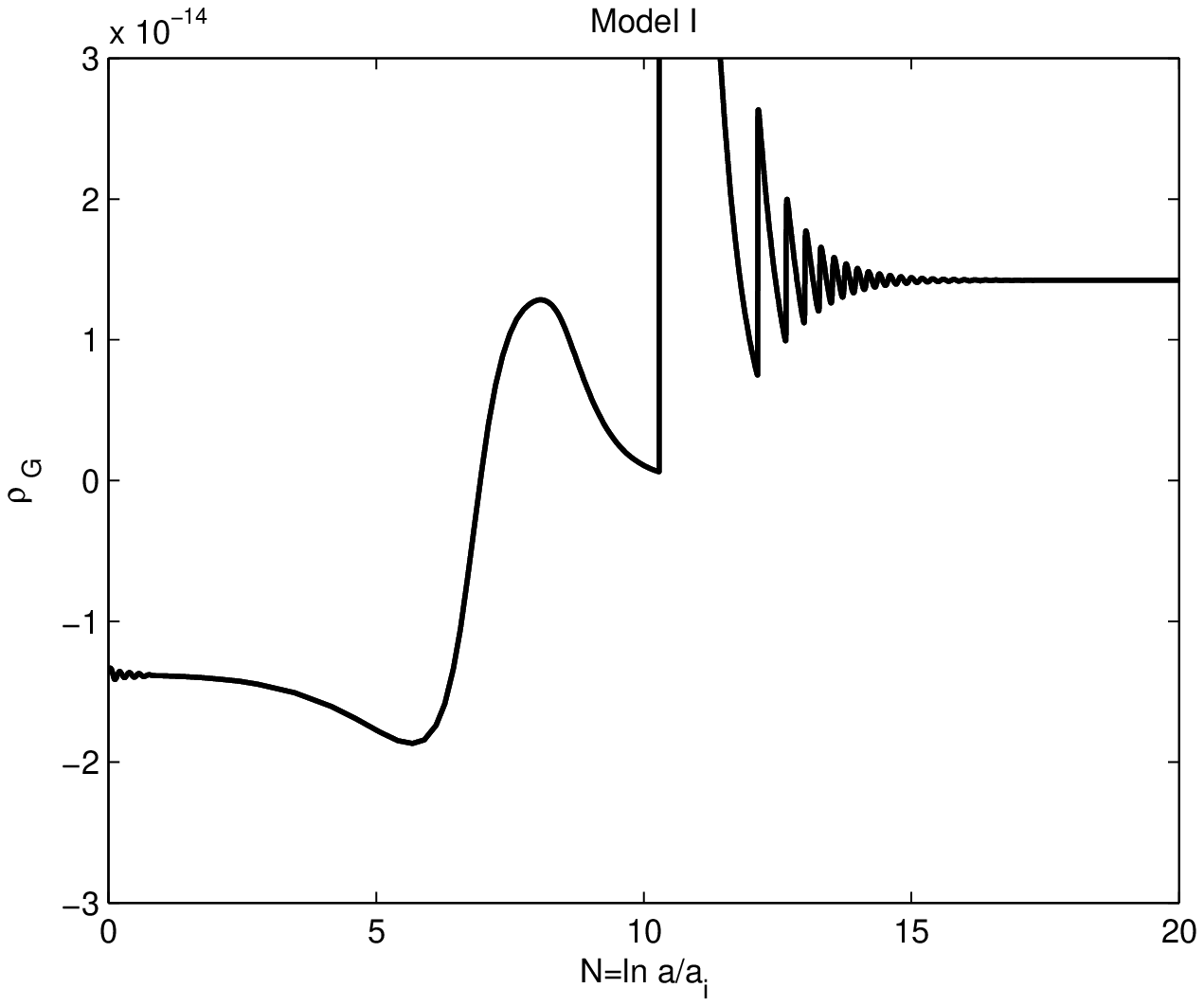}
      \vspace{4.0cm}
      \caption[]{The evolution of $\rho_{\rm G}$, Eq. (\ref{ro G1}), versus $N$ for model I. Auxiliary parameters and initial values as in Fig. \ref{JCAP2}.}
         \label{JCAP9}
   \end{figure}

\begin{figure}
\includegraphics{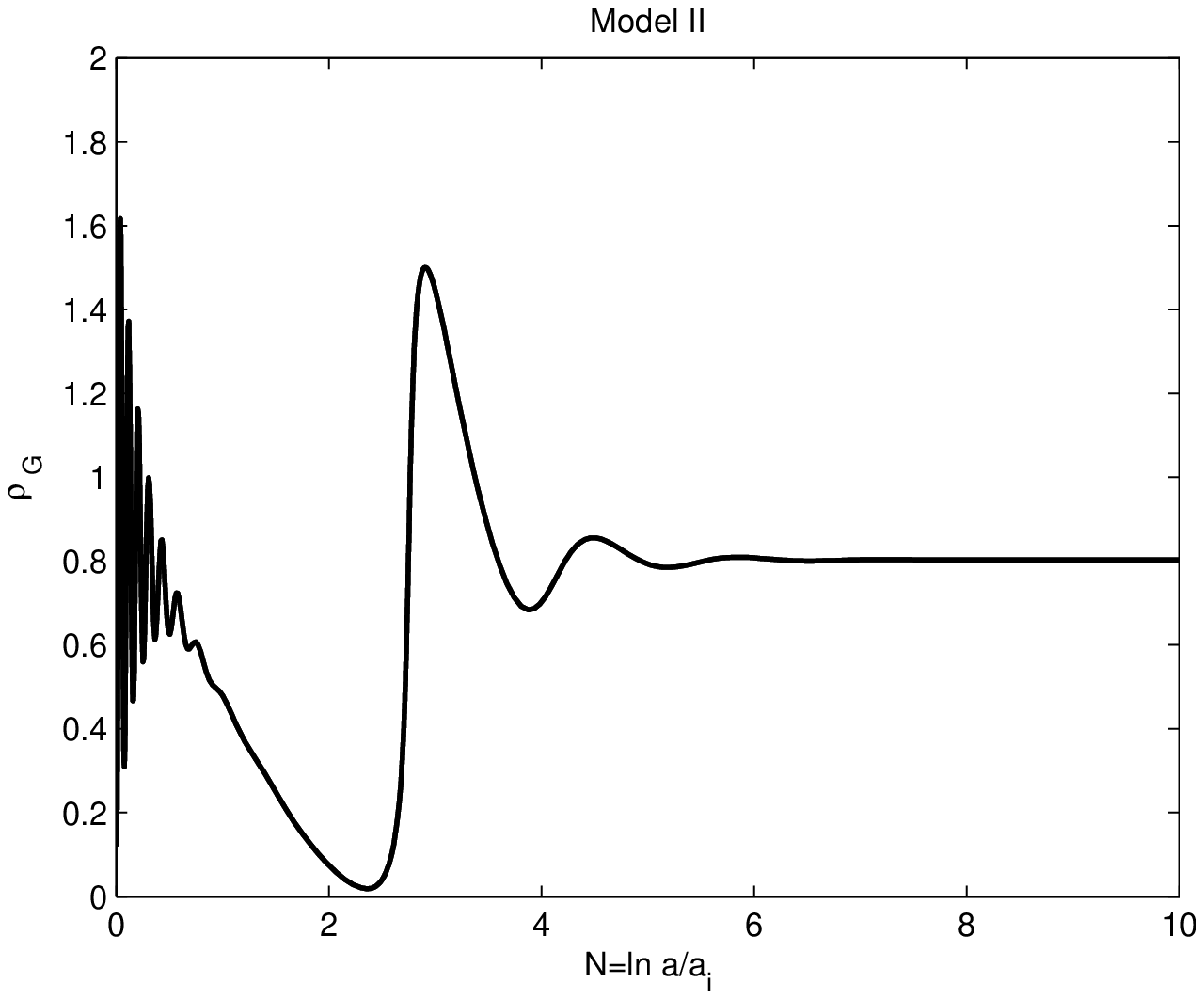}
      \vspace{4.0cm}
      \caption[]{Same as Fig. \ref{JCAP9} but for model II. Auxiliary parameters and initial values as in Fig. \ref{PLB2}.}
         \label{PLB9}
   \end{figure}

\begin{figure}
\includegraphics{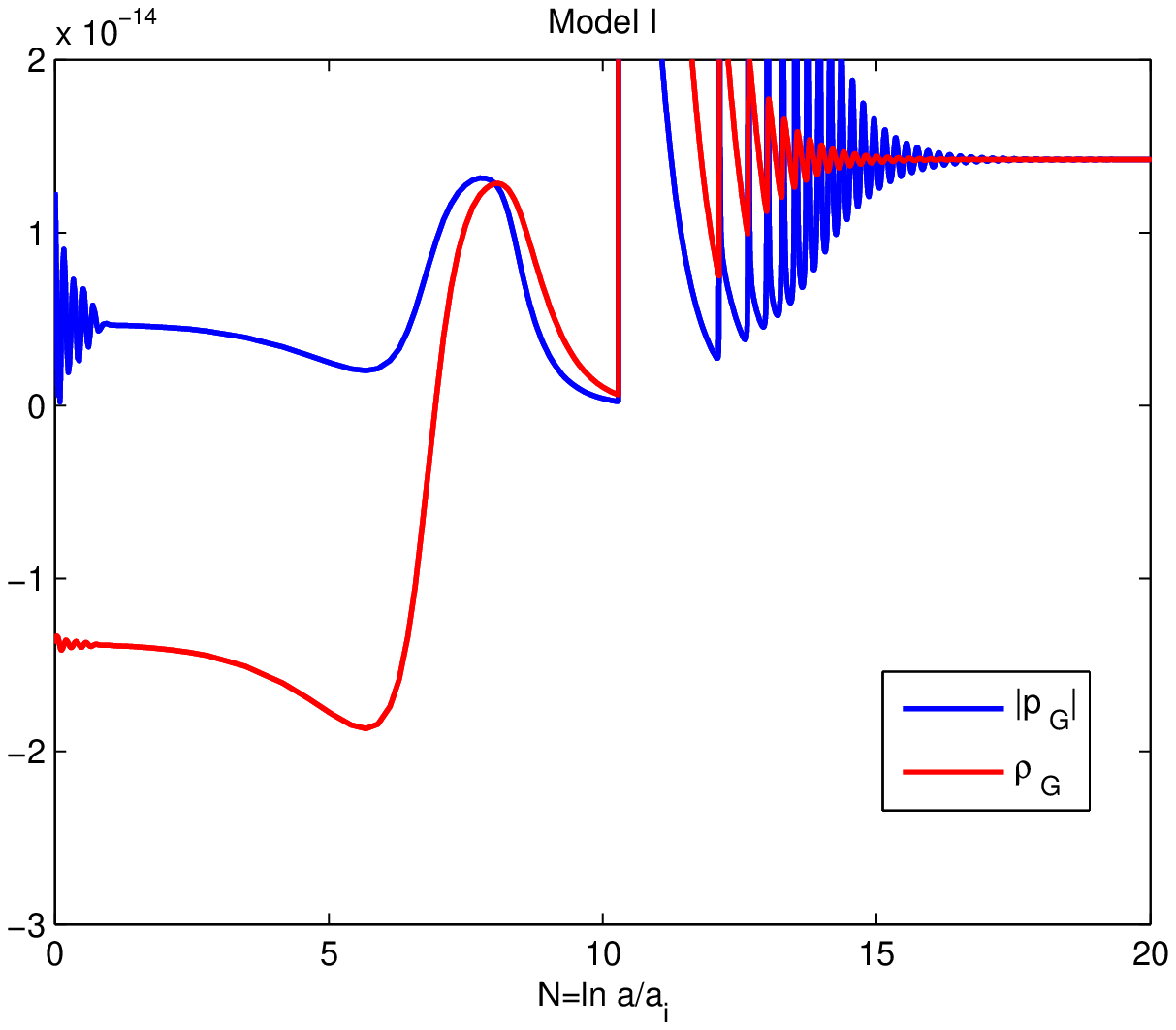}
      \vspace{4.0cm}
      \caption[]{The evolutions of $\rho_{\rm G}$ and $|p_{\rm G}|$, Eqs. (\ref{ro G1}) and (\ref{p G1}), versus $N$ for model I.
      Auxiliary parameters and initial values as in Fig. \ref{JCAP2}.}
         \label{JCAP10}
   \end{figure}

\begin{figure}
\includegraphics{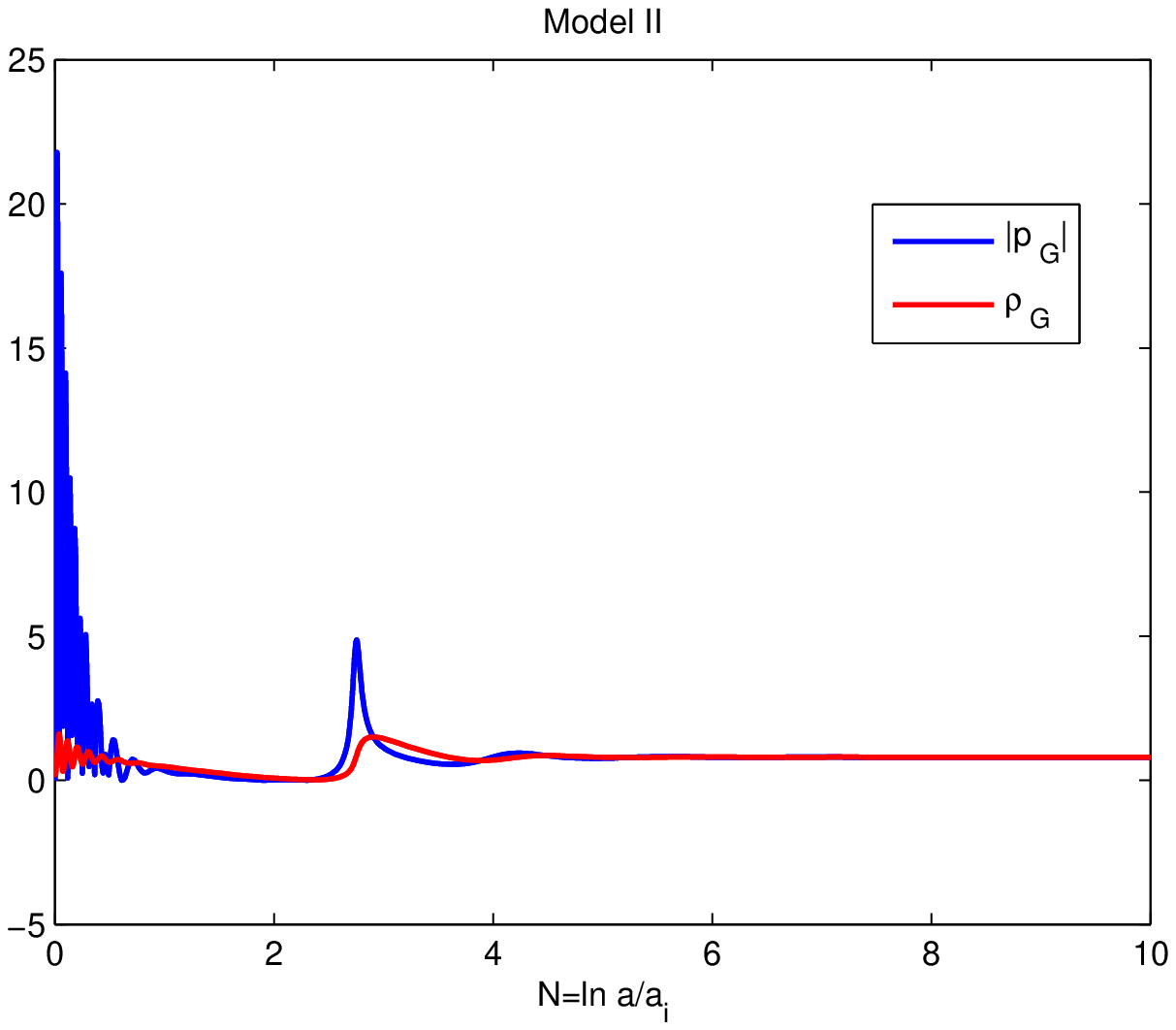}
      \vspace{4.0cm}
      \caption[]{Same as Fig. \ref{JCAP10} but for model II. Auxiliary parameters and initial values as in Fig. \ref{PLB2}.}
         \label{PLB10}
   \end{figure}

Using Eqs. (\ref{ro G1}) and (\ref{p G1}) the evolutionary behaviors
of $\rho_{\rm G}+p_{\rm G}$, $\rho_{\rm G}+3p_{\rm G}$, $\rho_{\rm
G}$ and $|p_{\rm G}|$ versus $N$ are plotted in Figs.
\ref{JCAP7}$-$\ref{PLB10}. Figures \ref{JCAP7}, \ref{JCAP8},
\ref{JCAP9} and \ref{JCAP10} illustrate that for model I, the energy
conditions during the standard radiation/matter dominated epochs are
violated in some ranges of $N$. Thereafter, $\rho_{\rm G}+p_{\rm
G}$, $\rho_{\rm G}+3p_{\rm G}$, $\rho_{\rm G}$ and $|p_{\rm G}|$
oscillate rapidly and finally the energy conditions, but the SEC,
for model I are satisfied in the late times. Figures \ref{PLB7},
\ref{PLB8}, \ref{PLB9} and \ref{PLB10} show that the energy
conditions for model II behave like model I. But the difference is
that in the future, $\rho_{\rm G}+p_{\rm G}$, $\rho_{\rm G}+3p_{\rm
G}$, $\rho_{\rm G}$ and $|p_{\rm G}|$ for model II vary smoothly
when the universe approaches a de Sitter regime.

\begin{figure}
\includegraphics{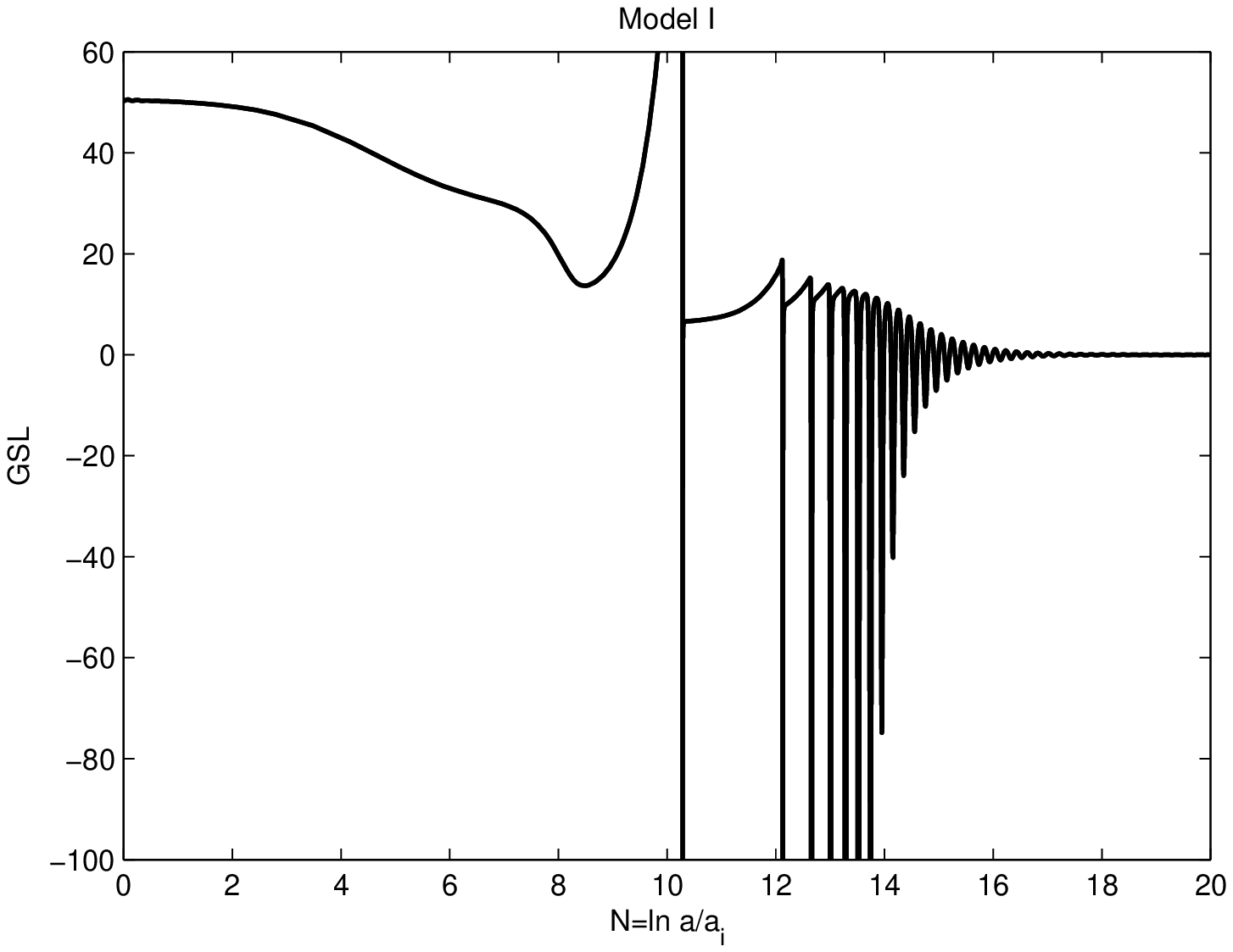}
      \vspace{4.0cm}
      \caption[]{The evolution of the GSL, Eq. (\ref{GSL 2}), versus $N$ for model I.
      Auxiliary parameters and initial values as in Fig. \ref{JCAP2}.}
         \label{JCAP6}
   \end{figure}

\begin{figure}
\includegraphics{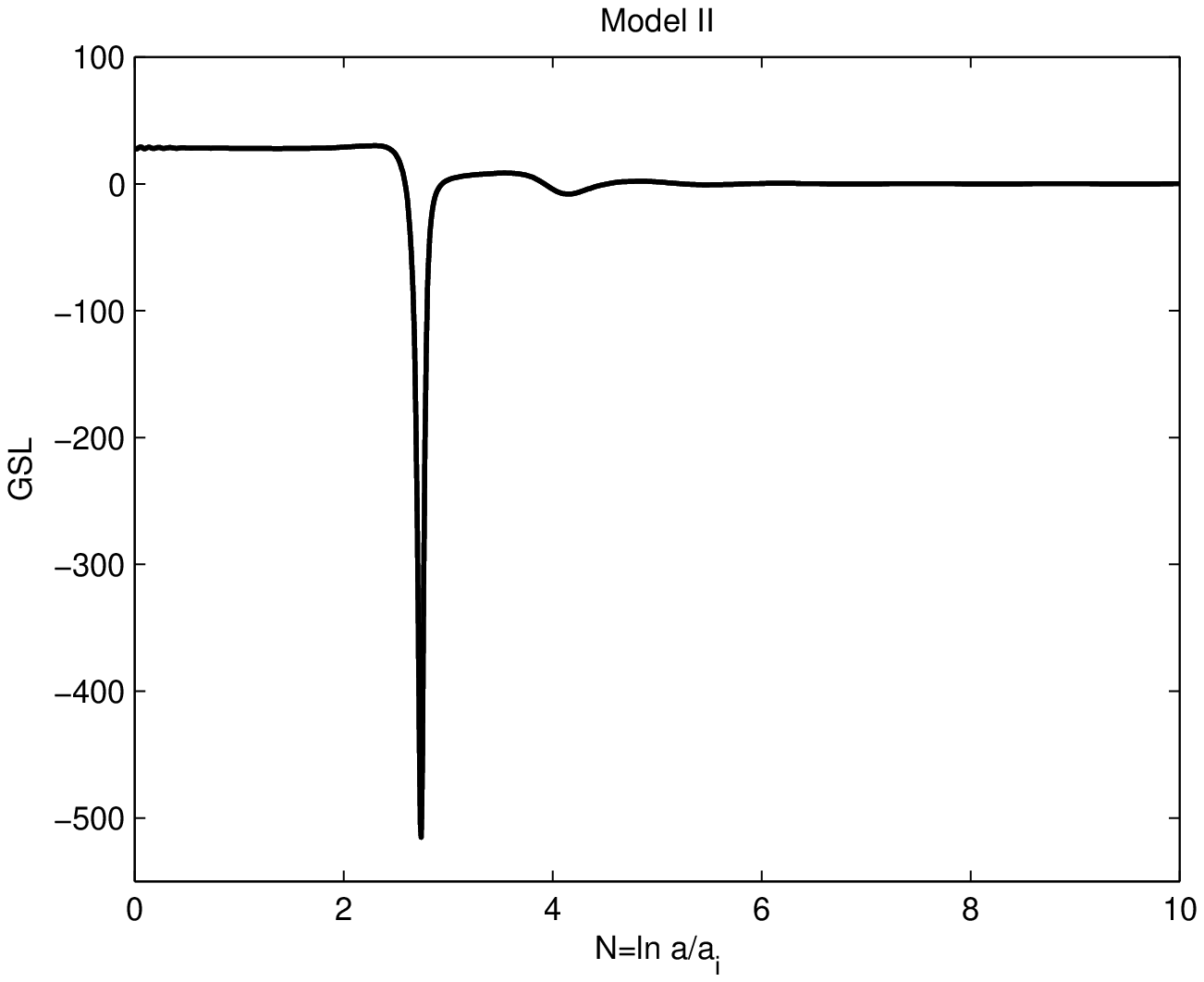}
      \vspace{4.0cm}
      \caption[]{Same as Fig. \ref{JCAP6} but for model II. Auxiliary parameters and initial values as in Fig. \ref{PLB2}.}
         \label{PLB6}
   \end{figure}

Finally, we examine the validity of the GSL for both models. In
Figs. \ref{JCAP6} and \ref{PLB6}, we plot the variation of the GSL,
Eq. (\ref{GSL 2}), versus $N$ for model I and model II,
respectively. Figures illustrate that for both models, the GSL
during the radiation/matter dominated epochs is fulfilled.
Thereafter when the universe enters the cosmic acceleration era,
i.e. $q<0$ see Figs. \ref{JCAP5} and \ref{PLB5}, the GSL does not
hold (i.e. $T_{\rm A}\dot{S}_{\rm tot}<0$) in some ranges of $N$. At
late times, the GSL for model I oscillates rapidly and for model II
approaches smoothly into the de Sitter universe, adiabatically (i.e.
$T_{A}\dot{S}_{\rm tot}=0$).

\section{Conclusions}

Here, we investigated the GSL of gravitational thermodynamics in the
framework of $f(G)$-gravity. To do so, we considered a spatially
flat FRW universe filled with the pressureless matter and radiation.
We supposed the boundary of the universe to be enclosed by the
dynamical apparent horizon with the Hawking radiation. We derived a
general relation for the GSL which its validity depends on
$f(G)$-model. Hence, for two viable $f(G)$-models containing
$f(G)=\alpha\left(G^{\frac{3}{4}}-\beta\right)^{\frac{2}{3}}$
\cite{JCAP} and
$f(G)=\lambda\frac{G}{\sqrt{G_{\ast}}}\arctan\left(\frac{G}{G_{\ast}}\right)
-\frac{\lambda}{2}\sqrt{G_{\ast}}~\ln\left(1+\frac{G^{2}}{G^{2}_{\ast}}\right)-\alpha\lambda\sqrt{G_{\ast}}$
\cite{PLB}, we first solved numerically the set of differential
equations governing the dynamics of $f(G)$-gravity. Consequently, we
obtained the evolutionary behaviors of the Hubble parameter, the
Gauss-Bonnet curvature invariant term, the pressureless matter,
radiation and DE density parameters, the effective EoS parameter and
the EoS parameter due to the $f(G)$ contribution as well as the
deceleration parameter. In addition, we turned to check the energy
conditions containing the NEC, SEC, WEC and DEC. Finally, we
examined the validity of the GSL for the two selected $f(G)$-models.
Our results show the following.

(i) The Hubble parameter $H$, the Gauss-Bonnet curvature invariant
term $|G|$ and the quantity $H^6f_{\rm GG}$ for both models at late
times go to a constant value when the universe enters a de Sitter
regime. Also both models have a stable de Sitter attractor, because
$H^6f_{\rm GG}$ satisfies the condition $0 < H_1^6 f_{\rm GG}(G_1) <
1/384$.

(ii) The density parameters $\Omega_{\rm r}$, $\Omega_{\rm m}$ and
$\Omega_{\rm G}$ for both models behave quite similar to those of
the $\Lambda$CDM model in the radiation and matter dominated epochs.

(iii) The EoS parameters $\omega_{\rm eff}$ and $\omega_{\rm G}$ for
both models start from the radiation/matter dominated epochs, then
enter the phantom region (i.e. $\omega<-1$) before reaching the de
Sitter attractor with $\omega=-1$.

(iv) The two selected $f(G)$ models can give rise to a late time
accelerated expansion phase of the universe. The deceleration
parameter for both models shows a cosmic deceleration $q>0$ to
acceleration $q<0$ transition which is compatible with the
observations \cite{Ishida}. Also for both models, $q$ is ended with
a stable de Sitter era (i.e. $q\rightarrow -1$).

(v) The NEC, SEC, WEC and DEC for both models during the
radiation/matter dominated eras are violated in some ranges of scale
factor. But in the late times when the universe approaches a de
Sitter regime, the all energy conditions, but the SEC, are
satisfied.

(vi) The GSL is fulfilled for both models during the standard
radiation/matter dominated epochs. But when the universe becomes
accelerating, the GSL is violated (i.e. $T_{\rm A}\dot{S}_{\rm
tot}<0$) in some ranges of scale factor. At late times, the
evolution of the GSL predicts an adiabatic behavior (i.e.
$T_{A}\dot{S}_{\rm tot}=0$) for the accelerated expansion of the
universe.

\section*{Acknowledgements}

The work of A. Abdolmaleki has been supported financially by Center
for Excellence in Astronomy \& Astrophysics of Iran
(CEAAI-RIAAM) under research project No. 1/3927.\\\\


\clearpage
\clearpage
\clearpage
\clearpage
\clearpage
\clearpage
\clearpage
\clearpage
\clearpage


\begin{thebibliography}{}

\bibitem{SN} M. Kowalski, et al. (Supernova Cosmology Project), Astrophys. J. {\bf 686}, 749 (2008).

\bibitem{LS} H. Lampeitl, et al., Mon. Not. R. Astron. Soc. {\bf 401}, 2331 (2010).

\bibitem{CMB1} E. Komatsu, et al. (WMAP Collaboration), Astrophys. J. Suppl. Ser. {\bf 192}, 18 (2011);
\\G. Hinshaw, et al. (WMAP Collaboration), Astrophys. J. Suppl. Ser. {\bf 208}, 19 (2013).

\bibitem{CMB2} P.A.R. Ade, et al. (Planck Collaboration), Astron. Astrophys. {\bf 571}, A16 (2014).

\bibitem{H} A.G. Riess, et al., Astrophys. J. {\bf 699}, 539 (2009).

\bibitem{Padmanabhan} T. Padmanabhan, Phys. Rep. {\bf 380}, 235 (2003);
\\P.J.E. Peebles, B. Ratra, Rev. Mod. Phys. {\bf 75}, 559 (2003);
\\C.G. Tsagas, A. Challinor, R. Maartens, Phys. Rep. {\bf 465}, 61
(2008);
\\M. Li, X.D. Li, S. Wang, Y. Wang, Commun. Theor. Phys. {\bf
56}, 525 (2011).

\bibitem{Sahni} V. Sahni, Class. Quantum Grav. {\bf 19}, 3435 (2002);
\\E.J. Copeland, M. Sami, S. Tsujikawa, Int. J. Mod. Phys. D {\bf 15}, 1753 (2006);
\\T. Padmanabhan, Gen. Relativ. Gravit. {\bf 40}, 529 (2008).

\bibitem{RoG} S. Capozziello, M.D. Laurentis, Phys. Rep. {\bf 509}, 167 (2011);
\\T. Clifton, P.G. Ferreira, A.Padilla, C. Skordis, Phys. Rep. {\bf 513}, 1 (2012).

\bibitem{fR} S. Nojiri, S.D. Odintsov, Phys. Rev. D {\bf 68}, 123512 (2003);
\\T.P. Sotiriou, V. Faraoni, Rev. Mod. Phys. {\bf 82}, 451 (2010);
\\S. Nojiri, S.D. Odintsov, Phys. Rep. {\bf 505}, 59 (2011);
\\K. Karami, M.S. Khaledian, JHEP {\bf 03}, 086 (2011).

\bibitem{fG} S. Nojiri, S.D. Odintsov, Phys. Lett. B {\bf 631}, 1 (2005).

\bibitem{fT} G.R. Bengochea, R. Ferraro, Phys. Rev. D {\bf 79}, 124019 (2009);
\\K. Karami, A. Abdolmaleki, S. Asadzadeh, Z. Safari, Eur. Phys. J. C {\bf 73}, 2565 (2013);
\\K. Karami, S. Asadzadeh, A. Abdolmaleki, Z. Safari, Phys. Rev. D {\bf 88}, 084034 (2013);
\\K. Karami, A. Abdolmaleki, Res. Astron. Astrophys. {\bf 13}, 757 (2013).

\bibitem{Sobouti} Y. Sobouti, Astron. Astrophys. {\bf 464}, 921 (2007).

\bibitem{Barrow0} B. Li, J.D. Barrow, D.F. Mota, Phys. Rev. D {\bf 76}, 044027 (2007);
\\S.C. Davis, Prog. Theor. Phys. Suppl. {\bf 172}, 81 (2008).

\bibitem{JCAP} S.Y. Zhou, E.J. Copeland, P.M. Saffin, JCAP {\bf 07}, 009 (2009).

\bibitem{PLB} A. De Felice, S. Tsujikawa, Phys. Lett. B {\bf 675}, 1 (2009).

\bibitem{Living} A. De Felice, S. Tsujikawa, Living Rev. Relativity {\bf 13}, 3
(2010).

\bibitem{Felic} A. De Felice, M. Hindmarsh, M. Trodden, JCAP {\bf 08}, 005
(2006).

\bibitem{de Sitter} G. Cognola, E. Elizalde, S. Nojiri, S. Odintsov, S. Zerbini, Phys. Rev.
D {\bf 75}, 086002 (2007);\\ S. Nojiri, S.D. Odintsov, P.V.
Tretyakov, Prog. Theor. Phys. Suppl. {\bf 172}, 81 (2008).


\bibitem{Wald} R.M. Wald, Phys. Rev. D {\bf 48}, 3427 (1993);
\\S.M. Carroll, et al., Phys. Rev. D {\bf 71}, 063513 (2005);
\\G. Cognola, E. Elizalde, S.Nojiri, S.D. Odintsov, S. Zerbini, JCAP {\bf 02}, 010 (2005).

\bibitem{AG} M. Alimohammadi, A. Ghalee, Phys. Rev. D {\bf 79}, 063006 (2009);
\\M. Alimohammadi, A. Ghalee, Phys. Rev. D {\bf 80}, 043006 (2009).

\bibitem{Metsaev} R.R. Metsaev, A.A. Tseytlin, Nucl. Phys. B {\bf 293}, 385 (1987).

\bibitem{BekHaw} J.D. Bekenstein, Phys. Rev. D {\bf 7}, 2333 (1973);
\\S.W. Hawking, Commun. Math. Phys. {\bf 43}, 199 (1975).

\bibitem{Jacobson} T. Jacobson, Phys. Rev. Lett. {\bf 75}, 1260 (1995).

\bibitem{Cai05} R.G. Cai, S.P. Kim, JHEP {\bf 02}, 050 (2005).

\bibitem{Akbar12} M. Akbar, R.G. Cai, Phys. Lett. B {\bf 635}, 7 (2006);
\\M. Akbar, R.G. Cai, Phys. Lett. B {\bf 648}, 243 (2007).

\bibitem{MK} R.X. Miao, M. Li, Y.G. Miao, JCAP {\bf 11}, 033 (2011).

\bibitem{Akbar} M. Akbar, R.G. Cai, Phys. Rev. D {\bf 75}, 084003
(2007).

\bibitem{Sheykhi1} A. Sheykhi, JCAP {\bf 05}, 019 (2009).

\bibitem{Izquierdo1} G. Izquierdo, D. Pav\'{o}n, Phys. Lett. B {\bf 639}, 1 (2006).

\bibitem{Sadjadi07} H. Mohseni Sadjadi, Phys. Rev. D {\bf 73}, 063525 (2006);
\\H. Mohseni Sadjadi, Phys. Rev. D {\bf 76}, 104024 (2007);
\\H. Mohseni Sadjadi, Phys. Lett. B {\bf 645}, 108 (2007).

\bibitem{Zhou07} J. Zhou, B. Wang, Y. Gong, E. Abdalla, Phys. Lett. B {\bf 652}, 86 (2007).

\bibitem{Gong07} Y. Gong, B. Wang, A. Wang, Phys. Rev. D {\bf 75}, 123516 (2007);
\\Y. Gong, B. Wang, A. Wang, JCAP {\bf 01}, 024 (2007).

\bibitem{Sheykhi2} A. Sheykhi, B. Wang, Phys. Lett. B {\bf 678}, 434 (2009);
\\A. Sheykhi, B. Wang, Mod. Phys. Lett. A {\bf 25}, 1199 (2010);
\\A. Sheykhi, Phys. Rev. D {\bf 81}, 104011 (2010);
\\A. Sheykhi, Eur. Phys. J. C {\bf 69}, 265 (2010);
\\A. Sheykhi, Class. Quantum Grav. {\bf 27}, 025007 (2010).

\bibitem{Karami1} K. Karami, JCAP {\bf 01}, 015 (2010);
\\K. Karami, S. Ghaffari, M.M. Soltanzadeh, Class. Quantum Grav. {\bf 27}, 205021 (2010);
\\K. Karami, A. Sheykhi, N. Sahraei, S. Ghaffari, Europhys. Lett. {\bf 93}, 29002 (2011);
\\K. Karami, A. Abdolmaleki, N. Sahraei, S. Ghaffari, JHEP {\bf 08}, 150 (2011).

\bibitem{Radicella} N. Radicella, D. Pav\'{o}n, Phys. Lett. B {\bf 691}, 121 (2010).

\bibitem{AA} K. Karami, A. Abdolmaleki, JCAP {\bf 04}, 007 (2012);
\\A. Abdolmaleki, T. Najafi, K. Karami, Phys. Rev. D {\bf 89}, 104041 (2014).

\bibitem{Geng} K. Bamba, C.Q. Geng, JCAP {\bf 11}, 008 (2011).



\bibitem{Capozziello2} S. Capozziello, V.F. Cardone, S. Carloni, A. Troisi, Int. J. Mod. Phys. D {\bf 12}, 1969 (2003);
\\S. Capozziello, V.F. Cardone, A. Troisi, Phys. Rev. D {\bf 71}, 043503 (2005).

\bibitem{Poisson} E. Poisson, W. Israel, Phys. Rev. D {\bf 41}, 1796 (1990);
\\S.A. Hayward, Phys. Rev. D {\bf 53}, 1938 (1996);
\\Y.G. Gong, A. Wang, Phys. Rev. Lett. {\bf 99}, 211301 (2007).

\bibitem{Cai09} R.G. Cai, L.M. Cao, Y.P. Hu, Class. Quantum Grav. {\bf 26}, 155018 (2009).



\bibitem{Halliwell} J.J. Halliwell, Phys. Lett. B {\bf 185}, 341 (1987);
\\E.J. Copeland, A.R. Liddle, D. Wands, Phys. Rev. D {\bf 57}, 4686 (1998);
\\S.Y. Zhou, Phys. Lett. B {\bf 660}, 7 (2008).

\bibitem{Ishida} E.E. Ishida, R.R.R Reis, A.V. Toribio, I. Waga, Astropart. Phys. {\bf 28}, 547 (2008).

\bibitem{HE} S.W. Hawking, G.F.R. Ellis, {\it The Large Scale Structure of Spacetime}, Cambridge University Press (1973);\\
R.M. Wald, {\it General Relativity}, The University of Chicago Press
(1984).


\bibitem{Bamba} K. Bamba, S. Nojiri, S.D. Odintsov, JCAP {\bf 10}, 045 (2008);\\
N.M. Garcia, T. Harko, F.S.N. Lobo, J.P. Mimoso, Phys. Rev. D {\bf
83}, 104032 (2011).

\bibitem{Barrow} J.D. Barrow, Class. Quantum Grav. {\bf 21}, L79
(2004);\\S. Nojiri, S.D. Odintsov, S. Tsujikawa, Phys. Rev. D {\bf
71}, 063004 (2005).


\end{thebibliography}
\end{document}